\documentclass[%
reprint,
superscriptaddress,
preprintnumbers,
amsmath,amssymb,
aps,
prd,
]{revtex4-2}

\usepackage{mathtools}
\usepackage{amsfonts}
\usepackage{mathrsfs}
\usepackage{bm}
\usepackage{bbold}
\usepackage{slashed}
\usepackage{dsfont}
\usepackage{float}
\usepackage{dcolumn}% Align table columns on decimal point
\usepackage{amssymb,amsmath}

\usepackage{graphicx}
\usepackage{subcaption}
\usepackage{color}
\usepackage{array}

\usepackage{placeins}
\usepackage{booktabs}
\usepackage{caption}
\usepackage{xspace}
\usepackage{hyperref}
\usepackage[nameinlink]{cleveref}
\usepackage{bookmark}

\newcommand{\Tr}{\ensuremath{\operatorname{Tr}}}

\newcolumntype{L}{>{\centering\arraybackslash}m{3cm}}

\definecolor{blue}{rgb}{0,0,1}

\definecolor{green}{rgb}{0,1,0}

\definecolor{red}{rgb}{1,0,0}

\definecolor{gray}{rgb}{.5,.5,.5}

\definecolor{darkgreen}{rgb}{.0,.5,.0}

\usepackage{xspace}
\usepackage{siunitx}
\usepackage{xfrac}
\usepackage{hyperref}
\usepackage[nameinlink]{cleveref}
\usepackage{appendix}
\usepackage{xifthen}
\usepackage{xcolor}
\hypersetup{
	colorlinks,
	linkcolor={red!75!black},
	citecolor={blue!75!black},
	urlcolor={blue!75!black}
}
\def\Eq#1{\Cref{#1}}

\def\eqref#1{\Cref{#1}}

\def\app#1{\hyperref[#1]{App.~\ref{#1}}}
\def\app#1{\Cref{#1}}

\def\0#1#2{\frac{#1}{#2}}

\graphicspath{{./figures/}{./}}

\begin{document}
%\preprint{}
	
\title{Functional renormalization group study of neutral and charged pion under magnetic fields in the quark-meson model}

\author{Rui Wen}
\email{rwen@ucas.ac.cn}
\affiliation{School of Nuclear Science and Technology, University of Chinese Academy of Sciences, Beijing, 100049,
  P.R. China}

\author{Shi Yin}
\affiliation{School of Physics, Dalian University of Technology, Dalian, 116024,
  P.R. China}
	
\author{Wei-jie Fu}
\affiliation{School of Physics, Dalian University of Technology, Dalian, 116024,
  P.R. China}
	
\author{Mei Huang}
\email{huangmei@ucas.edu.cn}
\affiliation{School of Nuclear Science and Technology, University of Chinese Academy of Sciences, Beijing, 100049,
  P.R. China}

	
\begin{abstract}

We calculated the masses of neutral and charged pion and pion decay constants under an extra magnetic field at zero temperature. The quantum fluctuations are integrated through the functional renormalization group. We consider the quark and meson propagators in the Landau level representation and weak-field expansion, respectively. The neutral pion mass monotonically decreases with the magnetic field, while the charged pion mass monotonically increases with the magnetic field. The pion decay constant and the quark mass show the magnetic catalysis behavior at vanishing temperature. The neutral pion mass and pion decay constant are quantitatively in agreement with the lattice QCD results in the region of $eB < 1.2 {\rm GeV}^2$, and no non-monotonic mass behavior for charged pion has been observed in this framework. 

\end{abstract}

%\pacs{11.30.Rd, %Chiral symmetries
%		11.10.Wx, %Finite-temperature field theory
%		05.10.Cc, %Renormalization group methods
%		12.38.Mh  %Quark-gluon plasma
%}                             % Classification Scheme.
\maketitle
%\tableofcontents

\section{Introduction}
\label{sec:int}

Studying Quantum Chromodynamics (QCD) matter under strong external magnetic field and vortical field have attracted many attentions in recent years. Relativistic heavy-ion collisions provide us a platform to study QCD matter under extreme conditions in the laboratory. In non-central heavy-ion collisions, the collision of two high-speed nuclei moving in opposite directions could create strong magnetic fields of order $\sim10^{18}$ Gauss \cite{Skokov:2009qp,Deng:2012pc}. Strong magnetic fields also exist in the early universe and magnetars \cite{Vachaspati:1991nm, Durrer:2013pga, Kiuchi:2015sga}. Understanding the strongly interacting matter in background magnetic fields requires a combination of the QCD and QED theories, which has brought about plenty of novel phenomena of magnetized quark matter, such as the chiral magnetic effect (CME) \cite{Kharzeev:2007jp, Kharzeev:2010gr}, magnetic catalysis (MC) \cite{Klevansky:1989vi,Klimenko:1990rh,Gusynin:1995nb}, inverse magnetic catalysis (IMC) \cite{Bali:2012zg, Tomiya:2019nym, Andersen:2021lnk}, diamagnetism at low temperature and paramagnetism at high temperature  \cite{Bali:2014kia}. These rich phenomena have attracted theoretical investigations in lattice Monte-Carlo simulations \cite{Bali:2012jv, Bali:2011qj, Bali:2017ian, Bignell:2020dze, Bornyakov:2013eya, Ding:2020hxw, Ding:2022tqn}, as well as model calculations, such as Nambu-Jona-Lasinio (NJL)
\cite{Inagaki:2003yi,Chao:2014wla,Yu:2014xoa, Coppola:2018vkw,Coppola:2019uyr,Fayazbakhsh:2014mca,Chaudhuri:2019lbw,Chaudhuri:2019lbw}, quark-meson (QM) model \cite{Ayala:2018zat, Kamikado:2013pya, Kamikado:2014bua} and AdS/QCD \cite{Mamo:2015dea,Li:2016gfn}, within mean-field approximation or functional methods \cite{Fukushima:2012xw, Braun:2014fua, Mueller:2015fka, Fu:2017vvg, Li:2019nzj}, see e.g., \cite{Shovkovy:2012zn, Andersen:2014xxa, Miransky:2015ava,Hattori:2023egw} for reviews. 

It is also valuable to study the meson spectrum of QCD under magnetic fields, which plays an important role in the understanding of the rich phenomena mentioned above. It is believed that the neutral pion is helpful to explain the inverse magnetic catalysis \cite{Fukushima:2012kc,Mao:2016fha}, and the charged pions can explain the diamagnetic around the pseudo-critical temperature \cite{Li:2019nzj}. The meson spectra have been widely studied in lattice QCD and effective models \cite{Bali:2011qj,Luschevskaya:2016epp,Bali:2017ian,Wang:2017vtn,Liu:2018zag,Ayala:2020dxs,Bignell:2020dze,Hidaka:2012mz,Li:2020hlp,Carlomagno:2022arc}. Recent Lattice calculation in \cite{Ding:2020hxw} showed that at zero temperature, the mass of the neutral $\pi$ meson decreases monotonously with the magnetic field, while that of the charged pions shows a non-monotonic behavior. Some efforts have been made to understand the pion mass behavior under magnetic field in low energy effective models \cite{Endrodi:2019whh,Xu:2020yag,Lin:2022ied,Xing:2021kbw,Kojo:2021gvm}. However, the mass behaviors of the neutral and charged pions under magnetic field have not been explained simultaneously. Besides, the lattice and effective model calculations are also extended to finite temperatures, see e.g., \cite{Ding:2022tqn,Sheng:2020hge,Mei:2022dkd, Mei:2020jzn, Islam:2023zyo}.

In this work, we employ the quark-meson model, which is also called the linear sigma model coupled to quarks (LSMq) \cite{Ayala:2018zat, Das:2019ehv} to calculate the meson masses and decay constants under a magnetic field. This model is well used to study the QCD phase diagrams \cite{Schaefer:2004en, Chen:2021iuo}, Equation of State (EoS) \cite{Schaefer:1999em, Herbst:2013ufa} as well as the fluctuations of conserved charges \cite{Wen:2018nkn, Fu:2021oaw}. Note that it can be transformed from the NJL model through a Hubbard-Stratonovich transformation \cite{Hubbard:1959ub, Jung:2016yxl}. The results of the mean-field approximation of the QM model coincide with the point-like particles. In this work, we include the quantum fluctuations through the functional renormalization group approach (FRG) \cite{Dupuis:2020fhh, Pawlowski:2005xe}, which is a functional continuum field approach.

This paper is organized as follows. In \Eq{sec:LEFT}, we introduce the low energy effective theory, i.e. the 2-flavor quark meson model. In \Eq{sec:flow_eq}, the choice of the regulator, propagators under a magnetic field are discussed and the flow equations are presented.In \Eq{sec:results}, we show the numerical results in our calculation, including the meson masses, quark masses and decay constants as functions of the strength of magnetic field.  In \Eq{sec:Vertexes}, we show the vertexes of the 2-flavor quark meson model.  In \Eq{sec:loopfunction}, the threshold functions of the flow equations are given.

%%%%%%%%%%%%%%%%%%%%%%%%%%%%%%%%%%%%%%%%%%%%%%%%%%%%%%%%%%%%%%%%%%%%%%%%%%%%%%%%%%%%%%%%%%%%%%%%%%%%%%%%%%%%%%%%%%%%%%%

\section{Low energy effective theory}
\label{sec:LEFT}

At high renormalization group (RG) scale, the first-principle QCD system only includes the degrees of freedom of quarks and gluons. As the RG scale decreases, due to the finite mass gap, the gluons are decoupled from the system, and their dynamics are integrated out,  left with gluonic background field and its potential.  Consequently, composite degrees of freedom, e.g., mesons and baryons, emerge naturally from the dynamics of elementary degrees of freedom, see, e.g., \cite{Fu:2019hdw, Braun:2014ata, Helmboldt:2014iya}. The degrees of freedom of the QCD system are transformed into those of quarks and hadrons, which can be described by low-energy effective models, such as the QM model and NJL model. 

The effective action of the two-flavor quark-meson model in Euclidean space reads \cite{Andersen:2013swa}
\begin{align}
\Gamma_k=&\int_x  \bar q \gamma_\mu (\partial_\mu-i Q A_\mu ) q +  \Tr(D_\mu \phi \cdot D_\mu \phi^\dagger) \nonumber \\
&+ h \bar q  (T^0 \sigma+i \gamma_5 \vec{T} \cdot \vec{\pi}) q +V_k (\rho)-c \sigma ,
\label{eq:effectiveaction_eq}
\end{align}
with $\int_x=\int d^4x$, $Q=diag(2/3,-1/3)e$ and $q=(u,d)^T$. Here, $\phi$ denotes the meson fields:
\begin{align}
\phi&=T^0 \sigma +\vec{T} \cdot \vec{\pi}=\frac{1}{2} \begin{pmatrix} \sigma+\pi^0 & \sqrt{2} \pi^+ \\
\sqrt{2} \pi^- & \sigma-\pi^0 \end{pmatrix} \,.
\end{align}
In \Eq{eq:effectiveaction_eq}, the potential $V(\rho)$ is chiral symmetric with $\rho \equiv \text{Tr}[\phi^\dagger\phi]=\frac{1}{2} (\sigma^2+\vec \pi^2)$, and $c \sigma$ is the linear sigma term, which explicitly breaks the chiral symmetry and accounts for the pion masses. The  covariant derivative of meson fields reads
\begin{align}
D_\mu \phi=\partial _\mu - i A_\mu [Q,\phi].
\end{align}
Without loss of generality, a homogeneous magnetic field of strength $B$ is assumed along the $z$-direction and the Landau gauge is adopted, i.e. $A_\mu =(0,0,xB,0)$. For convenience, we define $p_\perp=(p_1,p_2)$ and $p_\parallel=(p_0,p_3)$. 

The curvature masses are defined as the two-point correlation function at vanishing external momentum
\begin{align}
m_{\phi,\text{cur}}^2=\Gamma_{\phi\phi}^{(2)}(p_0=0,\vec{p}=0),
\end{align}
and for the $\pi$ and $\sigma$  meson, they are given as
\begin{align}
m_\pi^2=V'(\rho) \quad m_\sigma^2=V'(\rho) + 2 \rho V''(\rho).
\end{align}
The light quark mass is
\begin{align}
m_q=\frac{1}{2} h \sigma_0.
\end{align}
Here $\sigma_0$ is the vacuum expectation value of the sigma meson field, which is located at the minimum of the effective potential.
The mesonic decay constant is also related to the vacuum expectation value via:
\begin{align}
f_\pi=\sigma_0.\label{eq:fpi}
\end{align}
In this work, we employ the local potential approximation (LPA), which is the leading order of the derivative expansion. In other words, we ignore the mesonic and quark wave function renormalizations and the running of the Yukawa coupling. See \cite{Kamikado:2013pya} for a relevant discussion, where magnetic dependent wave function renormalizations beyond LPA are investigated in one-flavor case within the FRG approach.

\section{Flow equations and regulators}
\label{sec:flow_eq}

%
%%%%%%%%%%%%%%%%%%%%%%%%%%%%%
\begin{figure}[t]
\includegraphics[width=0.5\textwidth]{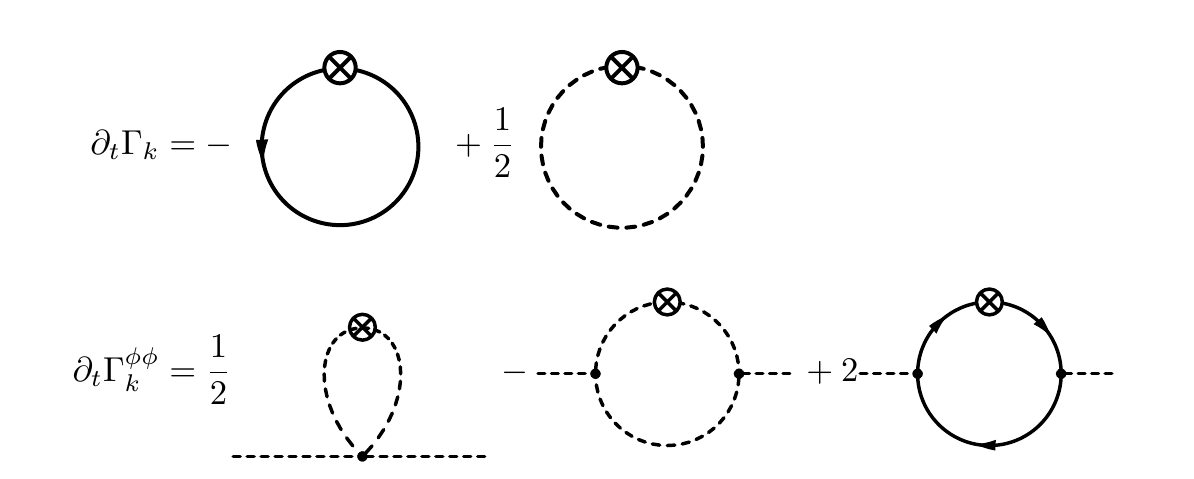}
\caption{Feynman diagrams of the flow equations for the effective potential (upper) and the mesonic two-point correlation functions (lower). The solid lines and dashed lines denote the quark and meson propagators, respectively. The crossed circles donates the infrared regulators, as shown in \Eq{eq:regulators}.}\label{flowEq}
\end{figure}
%%%%%%%%%%%%%%%%%%%%%%%%%%%%%
%

The evolution of the effective action with the RG scale is described by the Wetterich equation \cite{Wetterich:1992yh}, where an infrared (IR) cutoff scale $k$, i.e., the RG scale, is used to suppress quantum fluctuations of momenta below the scale. Starting from a high ultraviolet (UV) scale, say $\Lambda_{\mathrm{UV}}$, with the classical action as the initial condition, one is able to integrate-in quantum fluctuations of different modes successively by evolving the RG scale $k$ from UV to IR. The Wetterich equation for the effective action \Eq{eq:effectiveaction_eq} reads:
\begin{equation}
\partial_t \Gamma_k=\frac{1}{2} \text{Tr}[G_k^\phi(p) \partial_t R_k^B]-\text{Tr}[G_k^q(p) \partial_t R_k^F].\label{eq:Wetterich_eq}
\end{equation} 
Here $R_k$ denotes the regulators and $G^{\phi/q}_k(p)$ are scale-dependent propagators of mesons and quarks.

In the vacuum, the effective action satisfies the $O(4)$ space-time symmetry. When we consider an external magnetic field, the perpendicular (transverse) and parallel (longitudinal) directions to the magnetic field will split. Obviously, it will stay invariant in the temporal and $z$ directions at zero temperature. A commonly used $3d$ regulator for the spatial momenta breaks the $O(4)$ symmetry in the vacuum \cite{Yin:2019ebz}, while a regularization on the transverse momenta would give rise to non-physical artifacts \cite{Cao:2021rwx}. Therefore, in this work we adopt $2d$ regulators which regularize the temporal and longitudinal momenta, as follows
\begin{align}
R_k^B&=p_\parallel^2 r_B (p_\parallel^2/k^2), \nonumber \\
R_k^F&=i p_\parallel \cdot \gamma_\parallel r_F (p_\parallel^2/k^2),\label{eq:regulators}
\end{align} 
with $p_\parallel^2=p_0^2+p_3^2$ and the shape functions
\begin{align}
r_B(x)&=\bigg(\frac{1}{x}-1\bigg) \Theta (1-x) \nonumber \\
r_F(x)&=\bigg(\frac{1}{\sqrt{x}}-1 \bigg) \Theta(1-x).
\end{align} 
Here $\Theta(x)$ is the Heaviside step function. Notably, absence of regularization on the transverse momenta leads to a divergence for the flow equation of the potential $V_k$. Fortunately, the two-point correlation functions stay finite \cite{Kamikado:2013pya}. The summation of the Landau level can be calculated through the Hurwitz $\zeta$-function \cite{Andersen:2014xxa}
\begin{align}
\zeta(s,q)=\sum_n \frac{1}{(q+n)^s}.
\end{align}
In this work, we use a transverse momentum cutoff $\Lambda_{\perp}= 5\text{GeV}$ to calculate the  $u$-$d$ quark mixed threshold functions. We have checked that our results show no obvious dependence on the choices.

\subsection{propagators and flow equations}

The quark propagator in magnetic fields in the Schwinger scheme reads
\begin{align}
G(x,y)=e^{i \Phi(x_\perp,y_\perp)}\int\frac{d^4p}{(2 \pi)^4} e^{-i p (x-y)} \tilde{G}(p).
\end{align} 
Where prefactor $\Phi(x_\perp,y_\perp)=s_\perp (x^1+y^1)(x^2-y^2)|q_fB|/2$ with $s_{\perp} \equiv \text{sign}(q_fB)$ is the Schwinger phase \cite{Schwinger:1951nm}, which breaks the translational invariance. In this work, we ignore the Schwinger phase of the propagators under magnetic fields, and see, e.g., \cite{Coppola:2018vkw, Mao:2018dqe} for more discussions on the Schwinger phase with the Ritus scheme. Recently, it has been found that the Schwinger phase can be neglected when the meson masses are calculated \cite{Li:2020hlp}. The translationally invariant part of the quark propagator in the representation of Landau levels in the Euclidean space with the regulator reads \cite{Kamikado:2013pya, Andersen:2014xxa}:
\begin{align}
\tilde{G}_k^q(p)=\exp(-\frac{p_\perp^2}{|q_f B|})\sum_{n=0}^\infty \frac{(-1)^n D_n(p_{\parallel,R_F},p_\perp)}{p_{\parallel,R_F}^2 +2 n |q_f B| + m_f^2} \, ,
\end{align}
with $p_{\parallel,R_F} \equiv p_\parallel(1+r_F)$ and
\begin{align}
&D_n(p_{\parallel},p_\perp) \nonumber\\
=&(-i \gamma_\parallel p_\parallel +m_f)\Bigg[(1+i \gamma_1 \gamma_2 s_{\perp}) \mathcal{L}_n \bigg(\frac{2p_\perp^2}{|q_f B|}\bigg) \nonumber\\
&-(1-i \gamma_1 \gamma_2 s_{\perp})\mathcal{L}_{n-1}\bigg (\frac{2p_\perp^2}{|q_f B|}\bigg)\Bigg] \nonumber\\
&+4i \gamma_\perp p_\perp \mathcal{L}_{n-1}^{1}  \bigg (\frac{2p_\perp^2}{|q_f B|}\bigg).
\end{align} 
Here $\mathcal{L}_n^a(x)$ are the generalized Laguerre polynomials with $\mathcal{L}_{-1}^a(x)=0$. Similarly, the translationally invariant part of the scale-dependent meson propagator reads
\begin{align}
\tilde{G}_k^\phi(p)=&2 \exp(-\frac{p_\perp^2}{|q_\phi B|}) \nonumber \\
& \times\sum_{n=0}^{\infty} \frac{(-1)^n \mathcal{L}_n \big( \frac{2 p_\perp^2}{|q_\phi B|}\big)}{p_{\parallel,R_B}^2+(2n+1) |q_\phi B|+m_\phi^2}.
\end{align}
with $p_{\parallel,R_B} \equiv p_\parallel(1+r_B)^{\frac{1}{2}}$.

With the aforementioned setup, one is led to the flow equations of the effective potential:
\begin{align}
\partial_t V_k=&\frac{1}{2}\big[l_B(m_\sigma) +l_B(m_{\pi^0}) +2\,l_B(m_{\pi^\pm})\big] \nonumber \\
&- 4N_c \big[l_F(m_f,q_u)+l_F(m_f,q_d)\big].\label{flowV}
\end{align} 
The relevant Feynman diagrams are presented in the first line of \Eq{flowEq}. Here $l_B, l_F$ are threshold functions given in \Eq{sec:loopfunction}. By taking the second derivative of \Eq{eq:Wetterich_eq} with the pion fields, one arrives at the flow equation of two-point correlation function of the neutral pion as follows
\begin{align}
\partial_t \Gamma^{(2)}_{\pi^0\pi^0,k}=&\frac{1}{2}\big[V_{2 \pi^0 2\sigma} \mathcal{J}_B(\sigma) +V_{4 \pi^0} \mathcal{J}_B(\pi^0) \nonumber\\
&+2V_{2\pi^02\pi^\pm} \mathcal{J}_B(\pi^\pm)\big] -  V_{2 \pi^0 \sigma}^2\mathcal{J}_{2B}(\pi^0,\sigma) \nonumber \\
&+V^2_{\bar u u \pi^0}\mathcal{J}_F(u)+V^2_{\bar d d \pi^0}\mathcal{J}_F(d),\label{flow_neutralpion}
\end{align} 
and the flow equation of two-point correlation function of charged pions,
\begin{align}
\partial_t \Gamma^{(2)}_{\pi^\pm\pi^\pm,k}=&\frac{1}{2}\big[V_{2 \pi^\pm 2\sigma} \mathcal{J}_B(\sigma) +V_{2\pi^0\pi^\pm} \mathcal{J}_B(\pi^0) \nonumber\\
&+2V_{4\pi^\pm} \mathcal{J}_B(\pi^\pm)\big]-  V_{2 \pi^\pm \sigma}^2\mathcal{J}_{2B}(\pi^\pm,\sigma) \nonumber \\
&+V^2_{\bar u d \pi^\pm}\mathcal{J}_{2F}(u,d).
\label{flow_chargedpion}
\end{align}
Here $V_{[\cdots]}$ denote different vertices listed in \Eq{sec:Vertexes}, and $\mathcal{J}_B,\mathcal{J}_{2B},\mathcal{J}_{F},\mathcal{J}_{2F}$ are threshold functions, which are defined in \Eq{sec:loopfunction}. The corresponding Feynman diagrams are shown in the second line of \Eq{flowEq}. It can be readily verified that the neutral pion flow equation \Eq{flow_neutralpion} coincides with the flow equation of first order derivative of the potential, i.e.,
\begin{align}
\partial_t \Gamma^{(2)}_{\pi^0\pi^0,k}=\partial_tV_k'(\rho).
\end{align}

\subsection{weak-field expansion}
\label{subsec:weak_exp}

%
 %%%%%%%%%%%
\begin{table*}[t]
  \centering
  \begin{tabular}{cccc|cccc}
    \hline\hline
  $\lambda_1 [\text{MeV}]^2$ & $\lambda_2$ & $h$ & $c \,[\text{MeV}]^3$& $m_\pi\,[\text{MeV}]$&$m_\sigma\,[\text{MeV}]$  & $m_q\,[\text{MeV}]$  & $f_\pi\,[\text{MeV}]$ 
    \\ \hline
%    $(650)^2$  &6.0 & 6.4 & $1.75\times 10^6$ & 138 & 485 & 295 & 92\\
    $(740)^2$  &-5.0 & 6.4 & $4.5\times 10^6$ & 220 & 475 & 295 & 92 \\
    $(775)^2$  &6.0 & 6.4 & $1.6\times 10^7$ & 416 & 675 &295 & 92\\
    \hline\hline
  \end{tabular}
  \caption{Parameters for the initial conditions in \Eq{eq:initial,eq:effectiveaction_eq} and corresponding physical observables at $B=0$. If not mentioned explicitly, most of the results are calculated with  the parameters in the first line with $m_\pi=220$ MeV.}
  \label{tab:coeffs}
\end{table*}
%%%%%%%%%%%%
%

The number of Landau levels increases significantly in the region of small magnetic field. We do the computation in this region by utilizing the weak-field expansion method. The weak-field expansion for the quark propagator in the Euclidean space reads \cite{Chyi:1999fc, Ayala:2014gwa}
\begin{align}
&\tilde{G}_k^q(p) \nonumber \\
=&\frac{-i p_{\mu, R_F}\gamma_\mu+m_f}{p_{R_F}^2+m_f^2}+i\frac{\gamma_1\gamma_2(m_f- i \gamma_\parallel p_{\parallel,R_F})}{(p_{R_F}^2+m_f^2)^2} q_f B \nonumber\\
&+2 \frac{p_\perp^2(m_f-i\gamma_\parallel p _{\parallel,R_F})+i\gamma_\perp p_\perp(m_f^2+p^2_{\parallel,R_F})}{(p_{R_F}^2+m_f^2)^4}(q_f B)^2 \nonumber\\
&+\mathcal{O}(q_f B)^3.
\end{align}
Thus, one arrives at the quark loop function for the two-point correlation function of charged pions, as follows
\begin{align}
&\mathcal{J}_{2F}(u,d) 
=-\frac{k^4N_c}{2 \pi^2} \bigg[\frac{\Lambda_{\perp}^2}{(k^2+m_f^2)(k^2+m_f^2+\Lambda_{\perp}^2)}\nonumber\\
&+\Big(\frac{1}{4 (k^2+m_f^2)^3}+\frac{5 k^2+ 5 m_f^2+8\Lambda_{\perp}^2}{12(k^2+m_f^2+\Lambda_{\perp}^2)^4}\Big)(q_u B)(q_d B)\bigg] \nonumber\\
&+\mathcal{O}(B)^4.\label{eq:wk_chargedpion}
\end{align}
In the same way, the quark loops for the two-point correlation function of neutral pions read 
\begin{align}
&\mathcal{J}_{2F}(q_f)=-\frac{k^4N_c}{2 \pi^2} \bigg[\frac{\Lambda_{\perp}^2}{(k^2+m_f^2)(k^2+m_f^2+\Lambda_{\perp}^2)} \nonumber \\
&+\Big(\frac{1}{4 (k^2+m_f^2)^3}+\frac{5 k^2+ 5 m_f^2+8\Lambda_{\perp}^2}{12(k^2+m_f^2+\Lambda_{\perp}^2)^4}\Big)(q_f B)^2\bigg] \nonumber \\
&+\mathcal{O}(B)^4.\label{eq:wk_neutraldpion}
\end{align}
The weak-field expansion for the meson propagator reads \cite{Ayala:2004dx, Ayala:2018zat}
\begin{align}
\tilde{G}_k^\phi(p)=&\frac{1}{p_{R_B}^2+m_\phi^2}+\frac{p_\perp^2 -p_{\parallel,R_B}^2-m_\phi^2}{(p_{R_B}^2+m_\phi^2)^4}(q_\phi B)^2 \nonumber\\
&+ \mathcal{O}(q_\phi B)^4.
\end{align}
Then the weak-field expansions of the charged pion loop function $\mathcal{J}_B(\pi^\pm)$ and the pion-sigma loop function $\mathcal{J}_{2B}(\pi^\pm,\sigma)$ can be readily obtained, and their explicit expressions are listed in \Eq{sec:loopfunction}.

We find that the quark loops, as the last diagram in the second line of \Eq{flowEq} shows,  play the dominant role for the pion two-point correlation functions. When $\Lambda_{\perp} \rightarrow{\infty}$, the leading term in $B$ reads $1/(4 (k^2+m_f^2)^3) q_{f_1} q_{f_2} B^2$. For the charged pion, the signs of $q_u, q_d$ are opposite, so this term would make a negative contribution to the flow equation, implying that the contribution of quantum fluctuations to the charged pion mass is positive, which results in larger mass for charged pions in FRG than the point-like mass. On the contrary, for the neutral pion, the sign of $q_u^2$ or $q_d^2$ are positive. Consequently, the flow is increased and the mass of neutral pion is decreased in comparison to that in vacuum.

\section{numerical results}
\label{sec:results}

In this work, we solved the flow equation of effective potential by employing the Taylor expansion method around a fixed point, i.e.
\begin{align}
V_k(\rho)=\sum_n^{N_v} \frac{\lambda_{n,k}}{n!}(\rho-\kappa)^n.
\end{align} 
Here $\kappa$ denotes the expanding point, located at the minimal value of the effective potential with $k=0$. We choose the maximal order of the Taylor expansion to be $N_v=5$, and for more discussions on the convergence of the Taylor expansion see \cite{Pawlowski:2014zaa, Yin:2019ebz}. We have also checked the physical-point expanding method, in which the expanding point is the minimal value of the effective potential at every value of the RG scale $k$. We find that these two methods coincide with each other and produce consistent results. The UV cutoff is chosen to be $\Lambda_{\mathrm{UV}}=700$ MeV, where the initial condition of the effective potential reads
\begin{align}
V_{\mathrm{UV}}(\rho)=\lambda_1 \rho+ \frac{\lambda_2}{2} \rho^2.\label{eq:initial}
\end{align}
Here, the parameters of the initial conditions and the corresponding physical observables at $B=0$ are listed in \Eq{tab:coeffs}. In order to compare with the lattice QCD results, $m_\pi=220$ MeV and $m_\pi=416$ MeV are chosen. Note that if not mentioned explicitly, most of the results are calcluated with $m_\pi=220$ MeV.

%
%%%%%%%%%%%%%%%%%%%%%%%%%%%%%
\begin{figure}[t]
\includegraphics[width=0.5\textwidth]{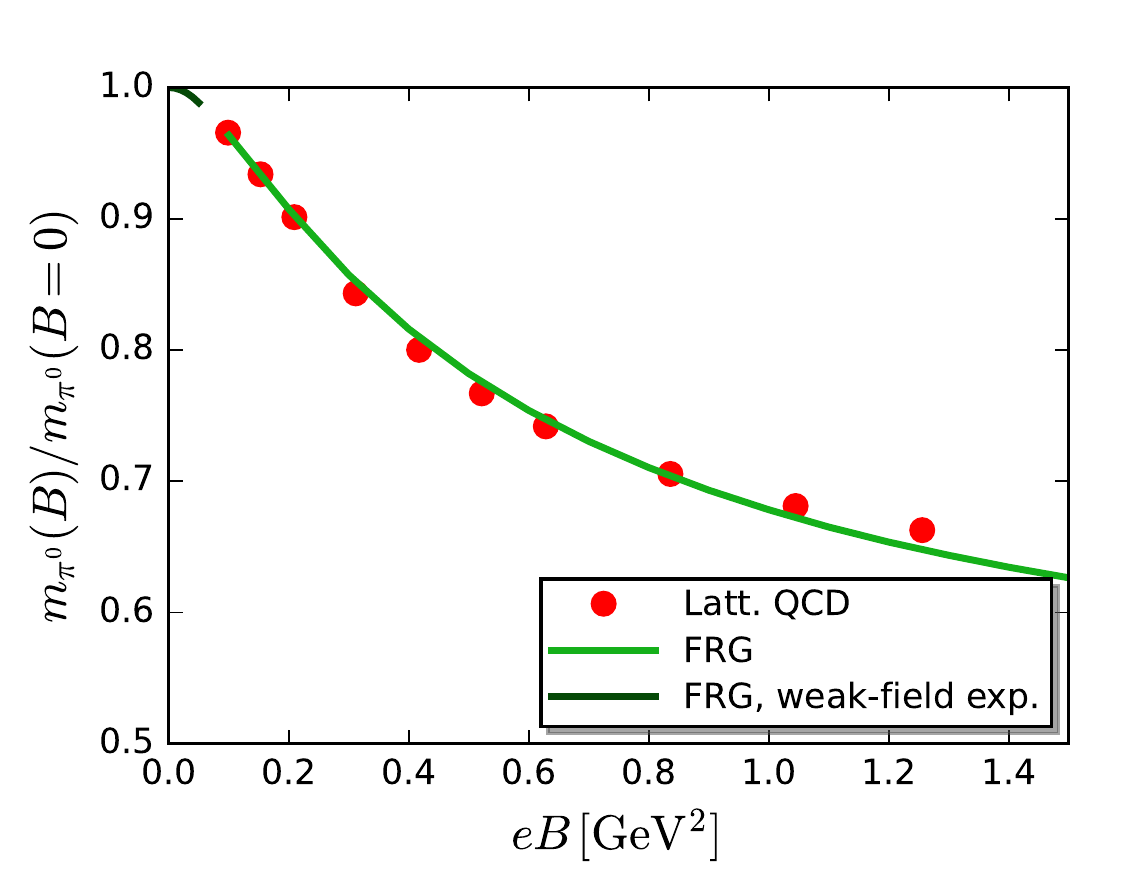}
\caption{Neutral pion mass $m_{\pi^0}$ as a function of the strength of magnetic field. The lattice QCD results are taken from Ref \cite{Ding:2020hxw}. }\label{pi0}
\end{figure}
%%%%%%%%%%%%%%%%%%%%%%%%%%%%%
%

In \Eq{pi0}, we show the neutral pion mass $m_{\pi^0}$ as a function of the strength of magnetic field  in comparison to the Lattice QCD results \cite{Ding:2020hxw}. In the region of small magnetic fields with $eB < 0.05 [\text{GeV}^2]$, we utilize the weak-field expansion method, while in other regions calculations are done through summation of the Landau levels. Our results are qualitatively or even quantitatively in agreement with the lattice results. If the neutral pion is regarded as a point particle, its masses will not change under magnetic fields. Due to the inner structure of the neutral pion, i.e. $\bar u u$ or $\bar d d$, the neutral pion mass decreases with the magnetic field, as discussed in \Eq{subsec:weak_exp}. The neutral pion mass decreases monotonically with the increase of magnetic fields, and the rate of decrease is gradually reduced. Finally, it tends to saturate in large magnetic fields. 

%
%%%%%%%%%%%%%%%%%%%%%%%%%%%%%
\begin{figure*}[t]
\includegraphics[width=1\textwidth]{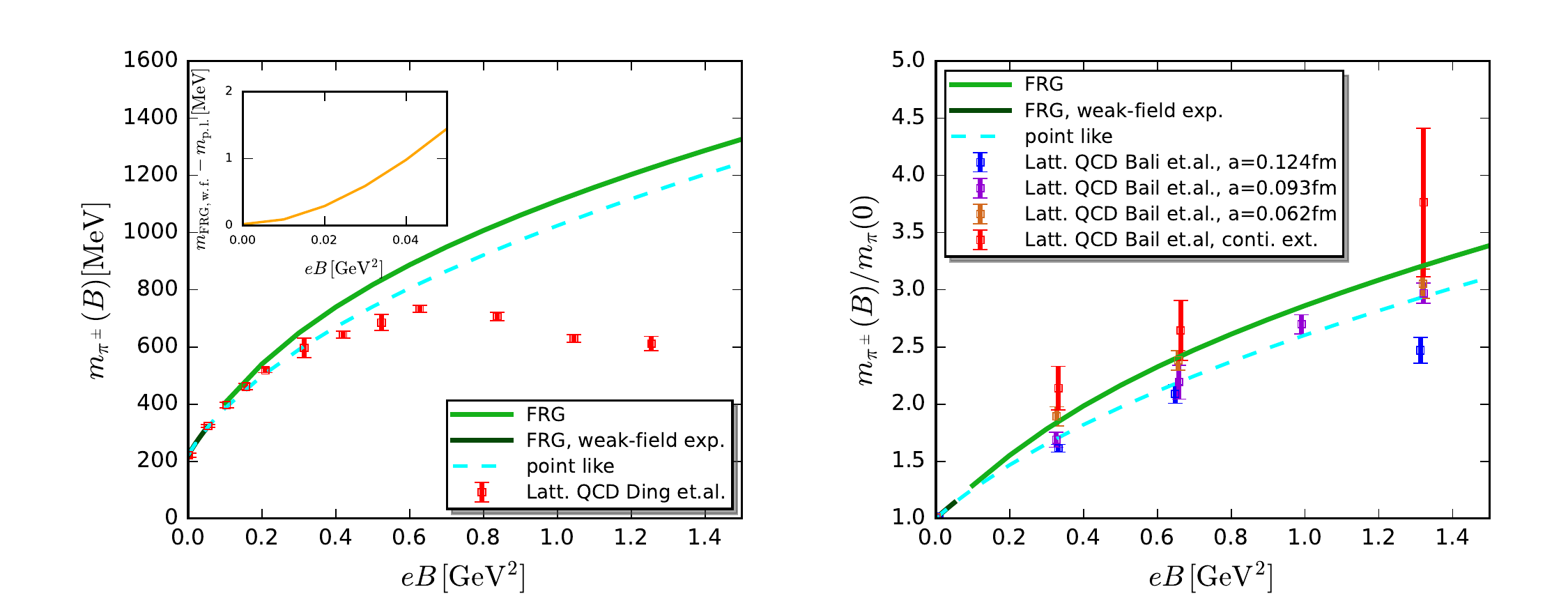}
\caption{Left panel: Charged pion mass $m_{\pi^\pm}$ as a function of the strength of magnetic field with $m_\pi(0)=220$ MeV. The lattice QCD results are constructed based on data from Ref \cite{Ding:2020hxw} and more details are shown in the text. In the inlay, we show the charged pion mass in the weak-field expansion with FRG subtracted by the point-like result. 
Right panel: Normalized charged pion mass $m_{\pi^\pm}(B)/m_{\pi }(0)$ as a function of magnetic fields with $m_\pi(0)=416$ MeV in comparison to  the relevant Lattice QCD \cite{Bali:2017ian}.}\label{pic}
\end{figure*}
%%%%%%%%%%%%%%%%%%%%%%%%%%%%%
%

The charged pion mass $m_{\pi^\pm}$ is defined as the lowest energy of quantum states for the charged pion \cite{Coppola:2018vkw}, i.e. $m_{\pi^\pm}(B)=E
_{\pi^\pm}|_{p_z=0,n=0}$. For the point particle of the charged pion,  the mass is given by $m_{\pi^\pm}(B)=\sqrt{m_{\pi^\pm}^2(B=0)+eB}$. According to the definition, we need to calculate the two-point correlation function $\Gamma_{\pi^\pm \pi^\pm}^{(2)}(p_\parallel=0, p_\perp=|eB|)$. Note, however, that it is challenging to integrate the loop functions $\mathcal{J}_{2 F} (u,d)$ and $\mathcal{J}_{2B}(\pi^\pm,\sigma)$ with finite external momenta. In our calculation, we use the approximation as follows
\begin{align}
m_{\pi^\pm}(B)=\sqrt{\Gamma_{\pi^\pm \pi^\pm}^{(2)}(p_\parallel=0, p_\perp=0)+eB}.
\end{align}
We also calculate $\Gamma_{\pi^\pm \pi^\pm}^{(2)}(p_\parallel=0, p_\perp=|eB|)$ at very large magnetic fields, and find that both results are consistent with each other.

In the left panel of \Eq{pic}, we plot the charged pion masses as functions of the strength of magnetic field with $m_\pi(B=0)=220$ MeV. In order to compare with the Lattice QCD results \cite{Ding:2020hxw},  where the computation is done with $m_{\pi}(B=0)\sim220 \text{MeV}$. We use lattice results of $m_{\pi^\pm}^2(B)-m_{\pi}^2(B=0)$ and construct
\begin{align}
m_{\pi^\pm}(B)=\sqrt{m_{\pi^\pm}^2(B)-m_{\pi}^2(B=0)+(220 \text{MeV})^2},
\end{align}
to be compared with FRG calculations. In the right panel of \Eq{pic}, we use the initial conditions in the second line in \Eq{tab:coeffs}, corresponding to $m_\pi(B=0)=416$ MeV, and compare the normalized charged pion mass $m_{\pi _\pm}(B)/m_{\pi }(0)$ with the lattice results with the same pion mass in the vacuum \cite{Bali:2017ian}. The charged pion masses in our calculation increase monotonically with the magnetic field. Our results are larger than the point-like charged pion masses and in agreement with the lattice QCD results in \cite{Bali:2017ian}. Similar results are also reported in the NJL calculation \cite{Coppola:2018vkw, Xu:2020yag}. However, for the lattice calculations in \cite{Ding:2020hxw}, the charged pion masses are smaller than the point-like results and exhibit nonmonotonic behaviors. This means our results receive an opposite contribution from the quantum fluctuation compared to the lattice QCD result in \cite{Ding:2020hxw}. The main contribution of quantum fluctuations comes from the $u$-$d$ quark loop, as discussed in the last paragraph of \Eq{subsec:weak_exp}, the leading order magnetic dependent quantum fluctuation of charged pion is opposite to that of the neutral pion, which could lead to the neutral pion masses smaller than point-like results and charged pion masses larger than point-like results in the region of weak magnetic field, as shown in the inlay in the left panel of \Eq{pic}. The calculation results with the Landau level representation coincide with those of weak-field expansion.  On the one hand, this discrepancy between the charged pion mass obtained in our calculations and that in lattice simulations in \cite{Ding:2020hxw} also probably arises from the approximations used in our calculations, such as neglect of the magnetic dependence of the Yukawa couplings and the wave function renormalizations. Our calculation is based on an effective model, which only contains the scalar and pseudoscalar channels, and other tensor structure channels and gluon dynamics are not taken into account \cite{Lin:2022ied}. On the other hand, the opposite quantum contribution could come from the lattice QCD calculation. Notably, the lattice cutoff in \cite{Ding:2020hxw} $a \simeq 0.117$ fm and no continuum limit is done, while in \cite{Bali:2017ian}  the continuum limit results are obtained, while the pion masses are much heavier than the physical value. Therefore, more detailed calculations and studies are required for both the lattice QCD and effective theories.

%
%%%%%%%%%%%%%%%%%%%%%%%%%%%%%
\begin{figure}[t]
\includegraphics[width=0.5\textwidth]{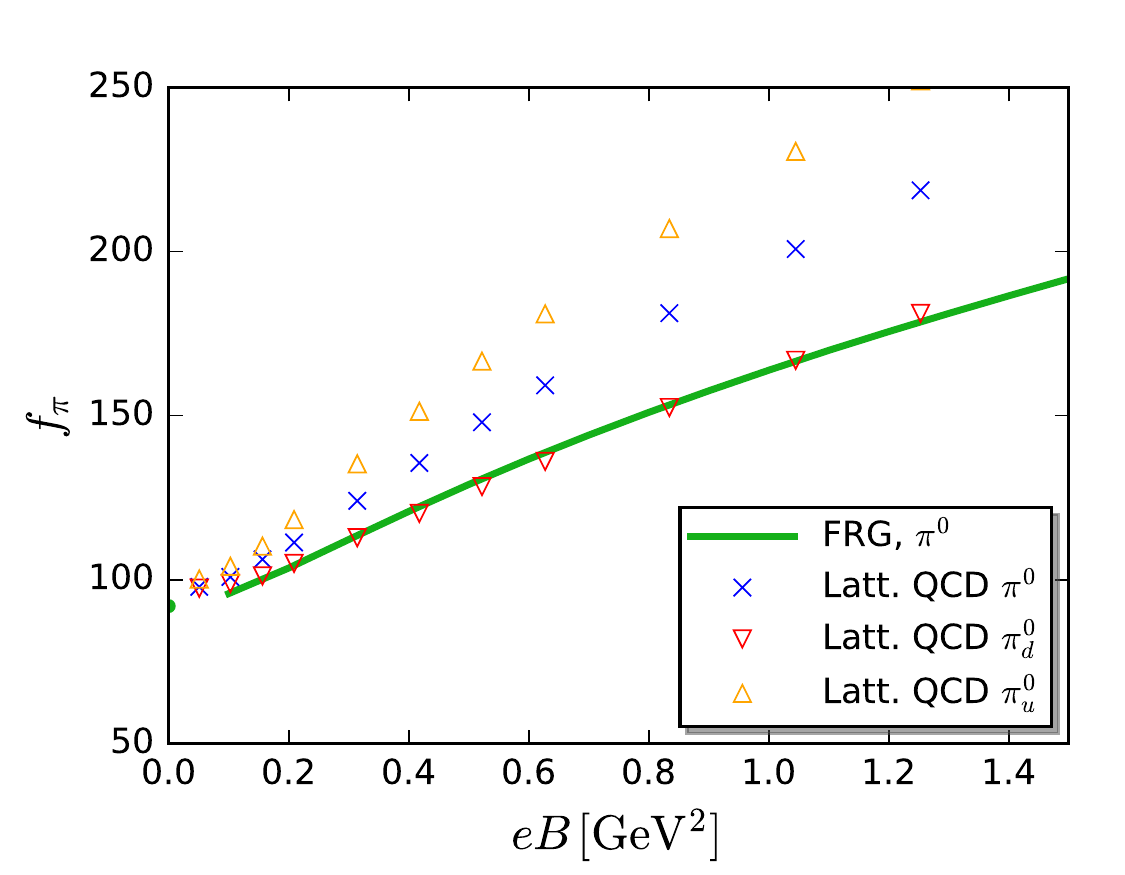}
\caption{Pion decay constant as a function of the strength of magnetic field. The lattice QCD results are taken from Ref \cite{Ding:2020hxw}. }\label{fpi}
\end{figure}
%%%%%%%%%%%%%%%%%%%%%%%%%%%%%
%

In \Eq{fpi}, we plot the pion decay constant as a function of the strength of magnetic field and compared it with the lattice QCD results \cite{Ding:2020hxw}.  For the 2-flavor QM model, the pion decay constant is determined by the minimum of the effective potential in \Eq{eq:fpi}. In the QM model, one cannot distinguish the $u$ or $d$ pion decay constants, and our results are close to that of $f_{\pi^0_d}$ in lattice QCD.

%
%%%%%%%%%%%%%%%%%%%%%%%%%%%%%
\begin{figure}[t]
\includegraphics[width=0.5\textwidth]{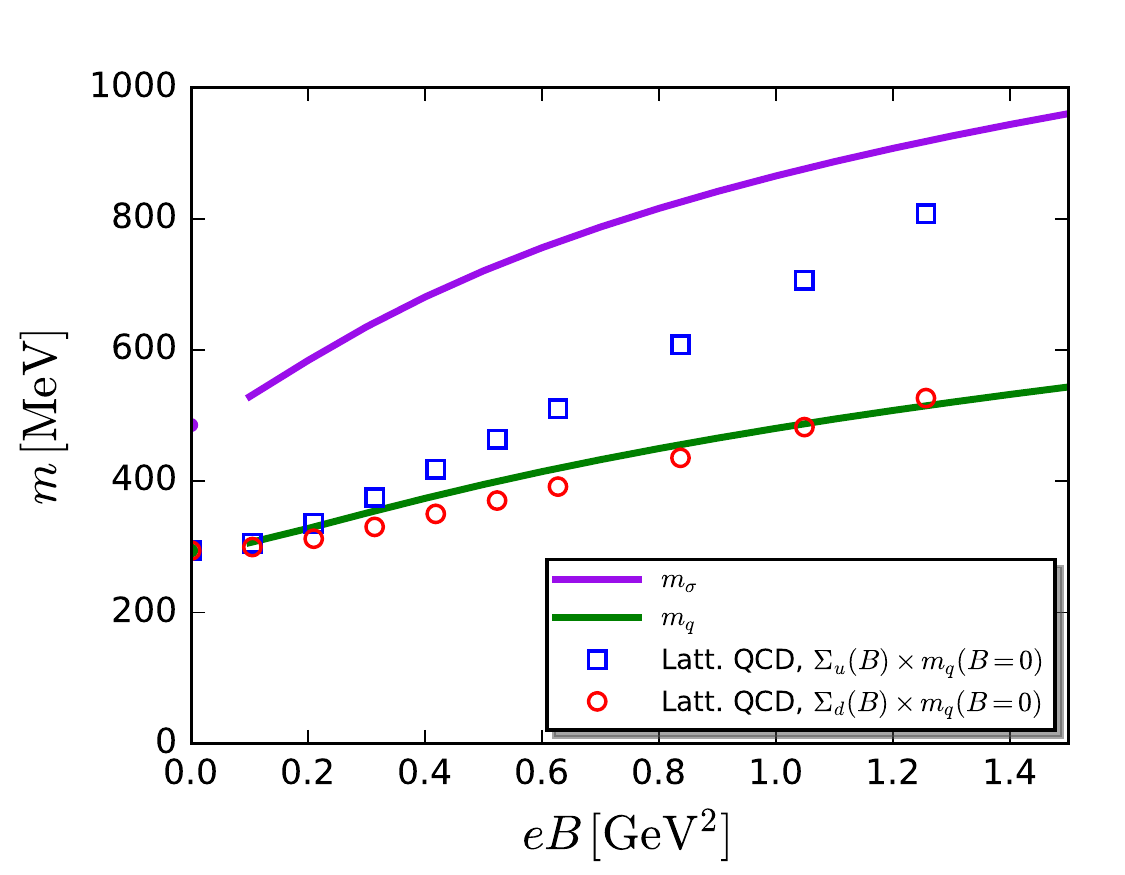}
\caption{Quark mass as a function of the strength of magnetic fields. The lattice QCD results are constructed from the quark chiral condensates in Ref \cite{Ding:2020hxw}. The $\sigma$ meson mass is also plotted. }\label{sigma}
\end{figure}
%%%%%%%%%%%%%%%%%%%%%%%%%%%%%
%

In \Eq{sigma}, we show the magnetic dependence of the sigma meson mass and light quark mass. The lattice QCD results are constructed from the quark chiral condensates in Ref \cite{Ding:2020hxw}. Similar to the pion decay constant $f_\pi$, the light quark mass is close to the $d$ quark results of the lattice QCD. Furthermore, due to the internal structure of mesons, the mass of sigma meson varies with the magnetic field. The sigma meson and light quark masses increase monotonically with the magnetic field. The decay constant, sigma meson mass, and light quark mass reflect chiral symmetry breaking increasing with magnetic fields, which is related to the magnetic catalysis.

\section{Conclusion}
\label{sec:sum}

This work calculates the meson masses and the pion decay constant at vanishing temperature under strong magnetic fields. The quantum fluctuations are successfully included using the FRG approach. The two-point correlation functions of neutral and charged pion are calculated. The neutral pion mass monotonically decreases with the magnetic field, while the sigma meson mass increases monotonically due to their internal structure. The decay constant and the light quark mass also increase with the magnetic field, reflecting the magnetic catalysis behavior at vanishing temperature. The neutral pion mass and pion decay constant are quantitatively in agreement with with the lattice QCD results especially in the range of $eB< 1.2 {\rm GeV^2}$. However, the charged pion mass is in agreement with the lattice results in \cite{Bali:2017ian} but no non-monotonic mass behavior for charged pion has been observed in this framework as shown in \cite{Ding:2020hxw}. This needs further investigation from both lattice QCD and functional methods.

It is noteworthy that this is our first preliminary attempt to calculate meson masses and the pion decay constant in the QM model under strong magnetic fields within the FRG approach, and there are many things we need to do in the future. In the upcoming work, we will go beyond the LPA truncation, which includes the magnetic dependent Yukawa couplings and wave function renormalizations, and calculate the spectral functions of the mesons. After that, we will extend them into finite temperatures and chemical potential. The strange quark and vector meson will also be included in future work.

\begin{acknowledgments}
We thank Chuang Huang, Jie Mei, Yang-yang Tan and Kun Xu for their valuable discussions. 
This work is supported in part by the National Natural Science Foundation of China (NSFC) Grant Nos: 12235016, 12221005, 12175030 and the Strategic Priority Research Program of Chinese Academy of Sciences under Grant No XDB34030000, the start-up funding from University of Chinese Academy of Sciences(UCAS), and the Fundamental Research Funds for the Central Universities.
\end{acknowledgments}

\appendix
 
\section{Vertexes}
\label{sec:Vertexes}

As mentioned above, we need the n-point vertexes to calculate the neutral and charged pion two-point correlation function \Eq{flow_neutralpion,flow_chargedpion}.
The n-point vertexes are defined as
\begin{align}
V_{\phi_1,\phi_2 \cdots \phi_n}=\frac{\partial^n\Gamma_k}{\partial  \phi_1 \partial  \phi_2 \cdots \partial  \phi_n}.
\end{align}
The quark-meson interaction in the 2-flavor QM model reads
\begin{align}
V_{\bar u d \pi^+}&=V_{\bar d u \pi^-}=\frac{\sqrt{2}}{2} h i \gamma_5\\
V_{\bar u u \pi^0}&=V_{\bar d d \pi^0}=\frac{1}{2} h i \gamma_5\\
V_{\bar u u \sigma}&=V_{\bar d d \sigma}=\frac{1}{2} h .
\end{align}
The nonvanishing mesonic three-point and four-point vertexes are
\begin{align}
V_{2 \pi^\pm \sigma}&=V_{2 \pi^0 \sigma}=\sigma V ''(\rho)\\
V_{3 \sigma} &= 3 \sigma V''(\rho)+\sigma^3 V'''(\rho)\\
V_{2 \pi^\pm 2\sigma}&=V_{2 \pi^0 2\sigma}=V ''(\rho)+\sigma^2 V '''(\rho)\\
V_{2 \pi^\pm 2\pi^0}&=V ''(\rho) \\
V_{4 \pi^0}&= 3 V ''(\rho) \\
V_{4 \pi^\pm}&= 2 V ''(\rho).
\end{align}

\section{loop functions}
\label{sec:loopfunction}
The threshold functions of the effective potential for neutral meson $\pi^0,\sigma$ in \Eq{flowV} read
\begin{align}
l_B(m_\phi)=\frac{k^4}{16 \pi^2} ( \log(k^2+ m_\phi^2+\Lambda_{\perp}^2)-\log(k^2+ m_\phi^2)).
\end{align} 
For the charged pion under magnetic fields, the threshold function is
\begin{align}
l_B(m_\phi)=\frac{k^4}{8 \pi^2}|q_\phi B| \sum_{n=0}^{\Lambda_{\perp,n}}\frac{1}{k^2+m_\phi^2+(2n+1) |q_\phi B|}.
\end{align} 
The quark loop function for the effective potential in the vacuum:
\begin{align}
l_F=\frac{k^4}{16 \pi^2}\Big(\log(k^2+m_f^2+\Lambda_{\perp}^2)-\log(k^2+m_f^2)\Big),
\end{align}
and the quark loop threshold function under magnetic fields reads
\begin{align}
l_F=\frac{k^4}{16 \pi^2}|q_f B|\sum_{n=0}^{\Lambda_{\perp,n}}\sum_{s=\pm 1}\frac{1}{k^2+m_f^2+ |q_f B|(2n+1+s)}.
\end{align}

The loop function of the tadpole diagram in weak-field expansion reads
\begin{align}
\mathcal{J}_{B}(\phi)&=\frac{k^4}{8 \pi^2} \Bigg[-\frac{\Lambda_{\perp}^2}{(k^2+m_\phi^2)(k^2+m_\phi^2+\Lambda_{\perp}^2)} \nonumber\\
&+\bigg( \frac{1}{3 (k^2+m_\phi^2)^3}-\frac{k^2+m_\phi^2-5\Lambda_\perp^2)}{3 (k^2+m_\phi^2+\Lambda_\perp^2)^4}\bigg)(q_\phi B)^2 \nonumber \\
&+ \mathcal{O}(q_\phi B)^4.
\end{align}
For the loop functions of neutral meson, we just need to set $q_\phi=0$. And  loop functions of the charge pion in Landau leval representation reads
\begin{align}
\mathcal{J}_{B}(\phi)=- \frac{k^4 |q_\phi B|}{4 \pi^2} \sum_{n=0}^{\Lambda_{\perp,n}}\frac{1}{k^2+(2n+1)|q_\phi B|+m_\phi^2}.
\end{align}

The $\sigma-\pi$ loop functions in weak-field expansion read
\begin{align}
&\mathcal{J}_{2B}(\pi,\sigma) \nonumber\\
=&\frac{k^4}{8 \pi^2}\Bigg[\bigg(\frac{1}{(k^2+m_\sigma^2+\Lambda_\perp^2)(k^2+m_\pi^2+\Lambda_\perp^2)} \nonumber\\
&-\frac{1}{(k^2+m_\sigma^2)(k^2+m_\pi^2)}\bigg) \nonumber\\
&-\int_0^{\Lambda_\perp^2}  \bigg(\frac{5p_\perp^2-3 m_\pi^2-3k^2}{(k^2+m_\pi^2+p_\perp^2)^5(k^2+m_\sigma^2+p_\perp^2)}\nonumber\\
&+\frac{p_\perp^2-k^2-m_\pi^2}{(k^2+m_\pi^2+p_\perp^2)^4(k^2+m_\sigma^2+p_\perp^2)^2}\bigg)d p_\perp^2(q_\pi B)^2\nonumber\\
&+\mathcal{O}(q_\pi B)^4
\end{align}
with neutral pion $q_{\pi^0}=0$ and  charged pion $q_{\pi^\pm}=\pm e$.

The $\sigma-\pi^\pm$ loop function in Landau leval representation reads
\begin{align}
\mathcal{J}_{2B}(\pi^\pm,\sigma)=&- \frac{k^4}{4\pi^2}\sum_{n=0} ^{\Lambda_{\perp,n}}\bigg( \frac{1}{(k^2+(2n+1)|eB|+m_\pi^2)^2} \nonumber\\
&\int_0^\infty \frac{e^{-y} \mathcal{L}_n(2y) dy}{y^2+(k^2+m_\sigma^2)/|eB|} \nonumber\\
&+\frac{1}{(k^2+(2n+1)|eB|+m_\pi^2)|eB|} \nonumber\\
& \int_0^\infty \frac{e^{-y} \mathcal{L}_n(2y) dy}{(y^2+(k^2+m_\sigma^2)/|eB|)^2}\bigg)
\end{align}

The weak-field expansion of quark loop of the two-point correction of pion have shown in \Eq{eq:wk_chargedpion} and \Eq{eq:wk_neutraldpion}. If we set $B=0$, they will come back to the representations of vacuum case. 

In Landau leval representation, the quark loop threshold functions of the neutral pion become
\begin{align}
\mathcal{J}_{2F}(q_f)=\frac{k^4 N_c}{2 \pi^2} \sum_{n=0}^{\Lambda_{\perp,n}}\sum_{s=\pm 1}\frac{1}{(k^2+m_f^2+ |q_f B|(2n+1+s))^2}
\end{align}
For the charged pion two-point correction, it contains a $u-d$ quark loop. The threshold function of this diagram reads
\begin{align}
\mathcal{J}_{2F}(u,d)=&\frac{ N_c k^4 B}{\pi^2} \sum_{n_1, n_2}^{\Lambda_{\perp,n} }(-1)^{(n_1+n_2)}[((\bar G^u_{n_1})^2 \bar G^d_{n_2}\nonumber\\
&+\bar G^u_{n_1}(\bar G^d_{n_2})^2)((k^2+m_f^2)(LL(n_1,n_2-1)\nonumber\\
&+LL(n_1-1,n_2))-8 B L^1L^1(n_1-1,n_2-1))\nonumber\\
&-\bar G^u_{n_1} \bar G^d_{n_2} (LL(n_1,n_2-1)+LL(n_1-1,n_2))
\end{align}
here 
\begin{align}
\bar G^{q_f}_{n}\equiv \frac{1}{k^2+m_f^2+ 2n|q_f B|)}
\end{align}
where we also define the $LL(n_1,n_2)$ and $L^1L^1(n_1,n_2)$ as the integrations of perpendicular direction
\begin{align}
LL(n_1,n_2)\equiv\int _0^\infty &dx \exp(-x(\frac{1}{|q_u|}+\frac{1}{|q_d|})) \\ \nonumber
&\mathcal{L}_{n_1} (\frac{2x}{|q_u|})\mathcal{L}_{n_2} (\frac{2x}{|q_d|})\\
L^1L^1(n_1,n_2)\equiv\int _0^\infty &dx x \exp(-x(\frac{1}{|q_u|}+\frac{1}{|q_d|})) \\ \nonumber
&\mathcal{L}^1_{n_1} (\frac{2x}{|q_u|})\mathcal{L}^1_{n_2} (\frac{2x}{|q_d|}),
\end{align}
with $\mathcal{L}_n^a(x)$ are the generalized Laguerre polynomials.
\bibliography{ref-lib}% Produces the bibliography via BibTeX.

\begin{thebibliography}{84}%
\makeatletter
\providecommand \@ifxundefined [1]{%
 \@ifx{#1\undefined}
}%
\providecommand \@ifnum [1]{%
 \ifnum #1\expandafter \@firstoftwo
 \else \expandafter \@secondoftwo
 \fi
}%
\providecommand \@ifx [1]{%
 \ifx #1\expandafter \@firstoftwo
 \else \expandafter \@secondoftwo
 \fi
}%
\providecommand \natexlab [1]{#1}%
\providecommand \enquote  [1]{``#1''}%
\providecommand \bibnamefont  [1]{#1}%
\providecommand \bibfnamefont [1]{#1}%
\providecommand \citenamefont [1]{#1}%
\providecommand \href@noop [0]{\@secondoftwo}%
\providecommand \href [0]{\begingroup \@sanitize@url \@href}%
\providecommand \@href[1]{\@@startlink{#1}\@@href}%
\providecommand \@@href[1]{\endgroup#1\@@endlink}%
\providecommand \@sanitize@url [0]{\catcode `\\12\catcode `\$12\catcode
  `\&12\catcode `\#12\catcode `\^12\catcode `\_12\catcode `\%12\relax}%
\providecommand \@@startlink[1]{}%
\providecommand \@@endlink[0]{}%
\providecommand \url  [0]{\begingroup\@sanitize@url \@url }%
\providecommand \@url [1]{\endgroup\@href {#1}{\urlprefix }}%
\providecommand \urlprefix  [0]{URL }%
\providecommand \Eprint [0]{\href }%
\providecommand \doibase [0]{https://doi.org/}%
\providecommand \selectlanguage [0]{\@gobble}%
\providecommand \bibinfo  [0]{\@secondoftwo}%
\providecommand \bibfield  [0]{\@secondoftwo}%
\providecommand \translation [1]{[#1]}%
\providecommand \BibitemOpen [0]{}%
\providecommand \bibitemStop [0]{}%
\providecommand \bibitemNoStop [0]{.\EOS\space}%
\providecommand \EOS [0]{\spacefactor3000\relax}%
\providecommand \BibitemShut  [1]{\csname bibitem#1\endcsname}%
\let\auto@bib@innerbib\@empty
%</preamble>
\bibitem [{\citenamefont {Skokov}\ \emph {et~al.}(2009)\citenamefont {Skokov},
  \citenamefont {Illarionov},\ and\ \citenamefont {Toneev}}]{Skokov:2009qp}%
  \BibitemOpen
  \bibfield  {author} {\bibinfo {author} {\bibfnamefont {V.}~\bibnamefont
  {Skokov}}, \bibinfo {author} {\bibfnamefont {A.~Y.}\ \bibnamefont
  {Illarionov}},\ and\ \bibinfo {author} {\bibfnamefont {V.}~\bibnamefont
  {Toneev}},\ }\bibfield  {title} {\bibinfo {title} {{Estimate of the magnetic
  field strength in heavy-ion collisions}},\ }\href
  {https://doi.org/10.1142/S0217751X09047570} {\bibfield  {journal} {\bibinfo
  {journal} {Int. J. Mod. Phys. A}\ }\textbf {\bibinfo {volume} {24}},\
  \bibinfo {pages} {5925} (\bibinfo {year} {2009})},\ \Eprint
  {https://arxiv.org/abs/0907.1396} {arXiv:0907.1396 [nucl-th]} \BibitemShut
  {NoStop}%
\bibitem [{\citenamefont {Deng}\ and\ \citenamefont
  {Huang}(2012)}]{Deng:2012pc}%
  \BibitemOpen
  \bibfield  {author} {\bibinfo {author} {\bibfnamefont {W.-T.}\ \bibnamefont
  {Deng}}\ and\ \bibinfo {author} {\bibfnamefont {X.-G.}\ \bibnamefont
  {Huang}},\ }\bibfield  {title} {\bibinfo {title} {{Event-by-event generation
  of electromagnetic fields in heavy-ion collisions}},\ }\href
  {https://doi.org/10.1103/PhysRevC.85.044907} {\bibfield  {journal} {\bibinfo
  {journal} {Phys. Rev. C}\ }\textbf {\bibinfo {volume} {85}},\ \bibinfo
  {pages} {044907} (\bibinfo {year} {2012})},\ \Eprint
  {https://arxiv.org/abs/1201.5108} {arXiv:1201.5108 [nucl-th]} \BibitemShut
  {NoStop}%
\bibitem [{\citenamefont {Vachaspati}(1991)}]{Vachaspati:1991nm}%
  \BibitemOpen
  \bibfield  {author} {\bibinfo {author} {\bibfnamefont {T.}~\bibnamefont
  {Vachaspati}},\ }\bibfield  {title} {\bibinfo {title} {{Magnetic fields from
  cosmological phase transitions}},\ }\href
  {https://doi.org/10.1016/0370-2693(91)90051-Q} {\bibfield  {journal}
  {\bibinfo  {journal} {Phys. Lett. B}\ }\textbf {\bibinfo {volume} {265}},\
  \bibinfo {pages} {258} (\bibinfo {year} {1991})}\BibitemShut {NoStop}%
\bibitem [{\citenamefont {Durrer}\ and\ \citenamefont
  {Neronov}(2013)}]{Durrer:2013pga}%
  \BibitemOpen
  \bibfield  {author} {\bibinfo {author} {\bibfnamefont {R.}~\bibnamefont
  {Durrer}}\ and\ \bibinfo {author} {\bibfnamefont {A.}~\bibnamefont
  {Neronov}},\ }\bibfield  {title} {\bibinfo {title} {{Cosmological Magnetic
  Fields: Their Generation, Evolution and Observation}},\ }\href
  {https://doi.org/10.1007/s00159-013-0062-7} {\bibfield  {journal} {\bibinfo
  {journal} {Astron. Astrophys. Rev.}\ }\textbf {\bibinfo {volume} {21}},\
  \bibinfo {pages} {62} (\bibinfo {year} {2013})},\ \Eprint
  {https://arxiv.org/abs/1303.7121} {arXiv:1303.7121 [astro-ph.CO]}
  \BibitemShut {NoStop}%
\bibitem [{\citenamefont {Kiuchi}\ \emph {et~al.}(2015)\citenamefont {Kiuchi},
  \citenamefont {Cerd\'a-Dur\'an}, \citenamefont {Kyutoku}, \citenamefont
  {Sekiguchi},\ and\ \citenamefont {Shibata}}]{Kiuchi:2015sga}%
  \BibitemOpen
  \bibfield  {author} {\bibinfo {author} {\bibfnamefont {K.}~\bibnamefont
  {Kiuchi}}, \bibinfo {author} {\bibfnamefont {P.}~\bibnamefont
  {Cerd\'a-Dur\'an}}, \bibinfo {author} {\bibfnamefont {K.}~\bibnamefont
  {Kyutoku}}, \bibinfo {author} {\bibfnamefont {Y.}~\bibnamefont {Sekiguchi}},\
  and\ \bibinfo {author} {\bibfnamefont {M.}~\bibnamefont {Shibata}},\
  }\bibfield  {title} {\bibinfo {title} {{Efficient magnetic-field
  amplification due to the Kelvin-Helmholtz instability in binary neutron star
  mergers}},\ }\href {https://doi.org/10.1103/PhysRevD.92.124034} {\bibfield
  {journal} {\bibinfo  {journal} {Phys. Rev. D}\ }\textbf {\bibinfo {volume}
  {92}},\ \bibinfo {pages} {124034} (\bibinfo {year} {2015})},\ \Eprint
  {https://arxiv.org/abs/1509.09205} {arXiv:1509.09205 [astro-ph.HE]}
  \BibitemShut {NoStop}%
\bibitem [{\citenamefont {Kharzeev}\ \emph {et~al.}(2008)\citenamefont
  {Kharzeev}, \citenamefont {McLerran},\ and\ \citenamefont
  {Warringa}}]{Kharzeev:2007jp}%
  \BibitemOpen
  \bibfield  {author} {\bibinfo {author} {\bibfnamefont {D.~E.}\ \bibnamefont
  {Kharzeev}}, \bibinfo {author} {\bibfnamefont {L.~D.}\ \bibnamefont
  {McLerran}},\ and\ \bibinfo {author} {\bibfnamefont {H.~J.}\ \bibnamefont
  {Warringa}},\ }\bibfield  {title} {\bibinfo {title} {{The Effects of
  topological charge change in heavy ion collisions: 'Event by event P and CP
  violation'}},\ }\href {https://doi.org/10.1016/j.nuclphysa.2008.02.298}
  {\bibfield  {journal} {\bibinfo  {journal} {Nucl. Phys. A}\ }\textbf
  {\bibinfo {volume} {803}},\ \bibinfo {pages} {227} (\bibinfo {year}
  {2008})},\ \Eprint {https://arxiv.org/abs/0711.0950} {arXiv:0711.0950
  [hep-ph]} \BibitemShut {NoStop}%
\bibitem [{\citenamefont {Kharzeev}\ and\ \citenamefont
  {Son}(2011)}]{Kharzeev:2010gr}%
  \BibitemOpen
  \bibfield  {author} {\bibinfo {author} {\bibfnamefont {D.~E.}\ \bibnamefont
  {Kharzeev}}\ and\ \bibinfo {author} {\bibfnamefont {D.~T.}\ \bibnamefont
  {Son}},\ }\bibfield  {title} {\bibinfo {title} {{Testing the chiral magnetic
  and chiral vortical effects in heavy ion collisions}},\ }\href
  {https://doi.org/10.1103/PhysRevLett.106.062301} {\bibfield  {journal}
  {\bibinfo  {journal} {Phys. Rev. Lett.}\ }\textbf {\bibinfo {volume} {106}},\
  \bibinfo {pages} {062301} (\bibinfo {year} {2011})},\ \Eprint
  {https://arxiv.org/abs/1010.0038} {arXiv:1010.0038 [hep-ph]} \BibitemShut
  {NoStop}%
\bibitem [{\citenamefont {Klevansky}\ and\ \citenamefont
  {Lemmer}(1989)}]{Klevansky:1989vi}%
  \BibitemOpen
  \bibfield  {author} {\bibinfo {author} {\bibfnamefont {S.~P.}\ \bibnamefont
  {Klevansky}}\ and\ \bibinfo {author} {\bibfnamefont {R.~H.}\ \bibnamefont
  {Lemmer}},\ }\bibfield  {title} {\bibinfo {title} {{Chiral symmetry
  restoration in the Nambu-Jona-Lasinio model with a constant electromagnetic
  field}},\ }\href {https://doi.org/10.1103/PhysRevD.39.3478} {\bibfield
  {journal} {\bibinfo  {journal} {Phys. Rev. D}\ }\textbf {\bibinfo {volume}
  {39}},\ \bibinfo {pages} {3478} (\bibinfo {year} {1989})}\BibitemShut
  {NoStop}%
\bibitem [{\citenamefont {Klimenko}(1991)}]{Klimenko:1990rh}%
  \BibitemOpen
  \bibfield  {author} {\bibinfo {author} {\bibfnamefont {K.~G.}\ \bibnamefont
  {Klimenko}},\ }\bibfield  {title} {\bibinfo {title} {{Three-dimensional
  Gross-Neveu model in an external magnetic field}},\ }\href
  {https://doi.org/10.1007/BF01015908} {\bibfield  {journal} {\bibinfo
  {journal} {Teor. Mat. Fiz.}\ }\textbf {\bibinfo {volume} {89}},\ \bibinfo
  {pages} {211} (\bibinfo {year} {1991})}\BibitemShut {NoStop}%
\bibitem [{\citenamefont {Gusynin}\ \emph {et~al.}(1996)\citenamefont
  {Gusynin}, \citenamefont {Miransky},\ and\ \citenamefont
  {Shovkovy}}]{Gusynin:1995nb}%
  \BibitemOpen
  \bibfield  {author} {\bibinfo {author} {\bibfnamefont {V.~P.}\ \bibnamefont
  {Gusynin}}, \bibinfo {author} {\bibfnamefont {V.~A.}\ \bibnamefont
  {Miransky}},\ and\ \bibinfo {author} {\bibfnamefont {I.~A.}\ \bibnamefont
  {Shovkovy}},\ }\bibfield  {title} {\bibinfo {title} {{Dimensional reduction
  and catalysis of dynamical symmetry breaking by a magnetic field}},\ }\href
  {https://doi.org/10.1016/0550-3213(96)00021-1} {\bibfield  {journal}
  {\bibinfo  {journal} {Nucl. Phys. B}\ }\textbf {\bibinfo {volume} {462}},\
  \bibinfo {pages} {249} (\bibinfo {year} {1996})},\ \Eprint
  {https://arxiv.org/abs/hep-ph/9509320} {arXiv:hep-ph/9509320} \BibitemShut
  {NoStop}%
\bibitem [{\citenamefont {Bali}\ \emph
  {et~al.}(2012{\natexlab{a}})\citenamefont {Bali}, \citenamefont {Bruckmann},
  \citenamefont {Endrodi}, \citenamefont {Fodor}, \citenamefont {Katz},\ and\
  \citenamefont {Schafer}}]{Bali:2012zg}%
  \BibitemOpen
  \bibfield  {author} {\bibinfo {author} {\bibfnamefont {G.~S.}\ \bibnamefont
  {Bali}}, \bibinfo {author} {\bibfnamefont {F.}~\bibnamefont {Bruckmann}},
  \bibinfo {author} {\bibfnamefont {G.}~\bibnamefont {Endrodi}}, \bibinfo
  {author} {\bibfnamefont {Z.}~\bibnamefont {Fodor}}, \bibinfo {author}
  {\bibfnamefont {S.~D.}\ \bibnamefont {Katz}},\ and\ \bibinfo {author}
  {\bibfnamefont {A.}~\bibnamefont {Schafer}},\ }\bibfield  {title} {\bibinfo
  {title} {{QCD quark condensate in external magnetic fields}},\ }\href
  {https://doi.org/10.1103/PhysRevD.86.071502} {\bibfield  {journal} {\bibinfo
  {journal} {Phys. Rev. D}\ }\textbf {\bibinfo {volume} {86}},\ \bibinfo
  {pages} {071502} (\bibinfo {year} {2012}{\natexlab{a}})},\ \Eprint
  {https://arxiv.org/abs/1206.4205} {arXiv:1206.4205 [hep-lat]} \BibitemShut
  {NoStop}%
\bibitem [{\citenamefont {Tomiya}\ \emph {et~al.}(2019)\citenamefont {Tomiya},
  \citenamefont {Ding}, \citenamefont {Wang}, \citenamefont {Zhang},
  \citenamefont {Mukherjee},\ and\ \citenamefont {Schmidt}}]{Tomiya:2019nym}%
  \BibitemOpen
  \bibfield  {author} {\bibinfo {author} {\bibfnamefont {A.}~\bibnamefont
  {Tomiya}}, \bibinfo {author} {\bibfnamefont {H.-T.}\ \bibnamefont {Ding}},
  \bibinfo {author} {\bibfnamefont {X.-D.}\ \bibnamefont {Wang}}, \bibinfo
  {author} {\bibfnamefont {Y.}~\bibnamefont {Zhang}}, \bibinfo {author}
  {\bibfnamefont {S.}~\bibnamefont {Mukherjee}},\ and\ \bibinfo {author}
  {\bibfnamefont {C.}~\bibnamefont {Schmidt}},\ }\bibfield  {title} {\bibinfo
  {title} {{Phase structure of three flavor QCD in external magnetic fields
  using HISQ fermions}},\ }\href {https://doi.org/10.22323/1.334.0163}
  {\bibfield  {journal} {\bibinfo  {journal} {PoS}\ }\textbf {\bibinfo {volume}
  {LATTICE2018}},\ \bibinfo {pages} {163} (\bibinfo {year} {2019})},\ \Eprint
  {https://arxiv.org/abs/1904.01276} {arXiv:1904.01276 [hep-lat]} \BibitemShut
  {NoStop}%
\bibitem [{\citenamefont {Andersen}(2021)}]{Andersen:2021lnk}%
  \BibitemOpen
  \bibfield  {author} {\bibinfo {author} {\bibfnamefont {J.~O.}\ \bibnamefont
  {Andersen}},\ }\bibfield  {title} {\bibinfo {title} {{QCD phase diagram in a
  constant magnetic background: Inverse magnetic catalysis: where models meet
  the lattice}},\ }\href {https://doi.org/10.1140/epja/s10050-021-00491-y}
  {\bibfield  {journal} {\bibinfo  {journal} {Eur. Phys. J. A}\ }\textbf
  {\bibinfo {volume} {57}},\ \bibinfo {pages} {189} (\bibinfo {year} {2021})},\
  \Eprint {https://arxiv.org/abs/2102.13165} {arXiv:2102.13165 [hep-ph]}
  \BibitemShut {NoStop}%
\bibitem [{\citenamefont {Bali}\ \emph {et~al.}(2014)\citenamefont {Bali},
  \citenamefont {Bruckmann}, \citenamefont {Endr\"odi}, \citenamefont {Katz},\
  and\ \citenamefont {Sch\"afer}}]{Bali:2014kia}%
  \BibitemOpen
  \bibfield  {author} {\bibinfo {author} {\bibfnamefont {G.~S.}\ \bibnamefont
  {Bali}}, \bibinfo {author} {\bibfnamefont {F.}~\bibnamefont {Bruckmann}},
  \bibinfo {author} {\bibfnamefont {G.}~\bibnamefont {Endr\"odi}}, \bibinfo
  {author} {\bibfnamefont {S.~D.}\ \bibnamefont {Katz}},\ and\ \bibinfo
  {author} {\bibfnamefont {A.}~\bibnamefont {Sch\"afer}},\ }\bibfield  {title}
  {\bibinfo {title} {{The QCD equation of state in background magnetic
  fields}},\ }\href {https://doi.org/10.1007/JHEP08(2014)177} {\bibfield
  {journal} {\bibinfo  {journal} {JHEP}\ }\textbf {\bibinfo {volume} {08}},\
  \bibinfo {pages} {177}},\ \Eprint {https://arxiv.org/abs/1406.0269}
  {arXiv:1406.0269 [hep-lat]} \BibitemShut {NoStop}%
\bibitem [{\citenamefont {Bali}\ \emph
  {et~al.}(2012{\natexlab{b}})\citenamefont {Bali}, \citenamefont {Bruckmann},
  \citenamefont {Constantinou}, \citenamefont {Costa}, \citenamefont {Endrodi},
  \citenamefont {Katz}, \citenamefont {Panagopoulos},\ and\ \citenamefont
  {Schafer}}]{Bali:2012jv}%
  \BibitemOpen
  \bibfield  {author} {\bibinfo {author} {\bibfnamefont {G.~S.}\ \bibnamefont
  {Bali}}, \bibinfo {author} {\bibfnamefont {F.}~\bibnamefont {Bruckmann}},
  \bibinfo {author} {\bibfnamefont {M.}~\bibnamefont {Constantinou}}, \bibinfo
  {author} {\bibfnamefont {M.}~\bibnamefont {Costa}}, \bibinfo {author}
  {\bibfnamefont {G.}~\bibnamefont {Endrodi}}, \bibinfo {author} {\bibfnamefont
  {S.~D.}\ \bibnamefont {Katz}}, \bibinfo {author} {\bibfnamefont
  {H.}~\bibnamefont {Panagopoulos}},\ and\ \bibinfo {author} {\bibfnamefont
  {A.}~\bibnamefont {Schafer}},\ }\bibfield  {title} {\bibinfo {title}
  {{Magnetic susceptibility of QCD at zero and at finite temperature from the
  lattice}},\ }\href {https://doi.org/10.1103/PhysRevD.86.094512} {\bibfield
  {journal} {\bibinfo  {journal} {Phys. Rev. D}\ }\textbf {\bibinfo {volume}
  {86}},\ \bibinfo {pages} {094512} (\bibinfo {year} {2012}{\natexlab{b}})},\
  \Eprint {https://arxiv.org/abs/1209.6015} {arXiv:1209.6015 [hep-lat]}
  \BibitemShut {NoStop}%
\bibitem [{\citenamefont {Bali}\ \emph
  {et~al.}(2012{\natexlab{c}})\citenamefont {Bali}, \citenamefont {Bruckmann},
  \citenamefont {Endrodi}, \citenamefont {Fodor}, \citenamefont {Katz},
  \citenamefont {Krieg}, \citenamefont {Schafer},\ and\ \citenamefont
  {Szabo}}]{Bali:2011qj}%
  \BibitemOpen
  \bibfield  {author} {\bibinfo {author} {\bibfnamefont {G.~S.}\ \bibnamefont
  {Bali}}, \bibinfo {author} {\bibfnamefont {F.}~\bibnamefont {Bruckmann}},
  \bibinfo {author} {\bibfnamefont {G.}~\bibnamefont {Endrodi}}, \bibinfo
  {author} {\bibfnamefont {Z.}~\bibnamefont {Fodor}}, \bibinfo {author}
  {\bibfnamefont {S.~D.}\ \bibnamefont {Katz}}, \bibinfo {author}
  {\bibfnamefont {S.}~\bibnamefont {Krieg}}, \bibinfo {author} {\bibfnamefont
  {A.}~\bibnamefont {Schafer}},\ and\ \bibinfo {author} {\bibfnamefont {K.~K.}\
  \bibnamefont {Szabo}},\ }\bibfield  {title} {\bibinfo {title} {{The QCD phase
  diagram for external magnetic fields}},\ }\href
  {https://doi.org/10.1007/JHEP02(2012)044} {\bibfield  {journal} {\bibinfo
  {journal} {JHEP}\ }\textbf {\bibinfo {volume} {02}},\ \bibinfo {pages}
  {044}},\ \Eprint {https://arxiv.org/abs/1111.4956} {arXiv:1111.4956
  [hep-lat]} \BibitemShut {NoStop}%
\bibitem [{\citenamefont {Bali}\ \emph {et~al.}(2018)\citenamefont {Bali},
  \citenamefont {Brandt}, \citenamefont {Endr\H{o}di},\ and\ \citenamefont
  {Gl\"a\ss{}le}}]{Bali:2017ian}%
  \BibitemOpen
  \bibfield  {author} {\bibinfo {author} {\bibfnamefont {G.~S.}\ \bibnamefont
  {Bali}}, \bibinfo {author} {\bibfnamefont {B.~B.}\ \bibnamefont {Brandt}},
  \bibinfo {author} {\bibfnamefont {G.}~\bibnamefont {Endr\H{o}di}},\ and\
  \bibinfo {author} {\bibfnamefont {B.}~\bibnamefont {Gl\"a\ss{}le}},\
  }\bibfield  {title} {\bibinfo {title} {{Meson masses in electromagnetic
  fields with Wilson fermions}},\ }\href
  {https://doi.org/10.1103/PhysRevD.97.034505} {\bibfield  {journal} {\bibinfo
  {journal} {Phys. Rev. D}\ }\textbf {\bibinfo {volume} {97}},\ \bibinfo
  {pages} {034505} (\bibinfo {year} {2018})},\ \Eprint
  {https://arxiv.org/abs/1707.05600} {arXiv:1707.05600 [hep-lat]} \BibitemShut
  {NoStop}%
\bibitem [{\citenamefont {Bignell}\ \emph {et~al.}(2020)\citenamefont
  {Bignell}, \citenamefont {Kamleh},\ and\ \citenamefont
  {Leinweber}}]{Bignell:2020dze}%
  \BibitemOpen
  \bibfield  {author} {\bibinfo {author} {\bibfnamefont {R.}~\bibnamefont
  {Bignell}}, \bibinfo {author} {\bibfnamefont {W.}~\bibnamefont {Kamleh}},\
  and\ \bibinfo {author} {\bibfnamefont {D.}~\bibnamefont {Leinweber}},\
  }\bibfield  {title} {\bibinfo {title} {{Pion magnetic polarisability using
  the background field method}},\ }\href
  {https://doi.org/10.1016/j.physletb.2020.135853} {\bibfield  {journal}
  {\bibinfo  {journal} {Phys. Lett. B}\ }\textbf {\bibinfo {volume} {811}},\
  \bibinfo {pages} {135853} (\bibinfo {year} {2020})},\ \Eprint
  {https://arxiv.org/abs/2005.10453} {arXiv:2005.10453 [hep-lat]} \BibitemShut
  {NoStop}%
\bibitem [{\citenamefont {Bornyakov}\ \emph {et~al.}(2014)\citenamefont
  {Bornyakov}, \citenamefont {Buividovich}, \citenamefont {Cundy},
  \citenamefont {Kochetkov},\ and\ \citenamefont
  {Sch\"afer}}]{Bornyakov:2013eya}%
  \BibitemOpen
  \bibfield  {author} {\bibinfo {author} {\bibfnamefont {V.~G.}\ \bibnamefont
  {Bornyakov}}, \bibinfo {author} {\bibfnamefont {P.~V.}\ \bibnamefont
  {Buividovich}}, \bibinfo {author} {\bibfnamefont {N.}~\bibnamefont {Cundy}},
  \bibinfo {author} {\bibfnamefont {O.~A.}\ \bibnamefont {Kochetkov}},\ and\
  \bibinfo {author} {\bibfnamefont {A.}~\bibnamefont {Sch\"afer}},\ }\bibfield
  {title} {\bibinfo {title} {{Deconfinement transition in two-flavor lattice
  QCD with dynamical overlap fermions in an external magnetic field}},\ }\href
  {https://doi.org/10.1103/PhysRevD.90.034501} {\bibfield  {journal} {\bibinfo
  {journal} {Phys. Rev. D}\ }\textbf {\bibinfo {volume} {90}},\ \bibinfo
  {pages} {034501} (\bibinfo {year} {2014})},\ \Eprint
  {https://arxiv.org/abs/1312.5628} {arXiv:1312.5628 [hep-lat]} \BibitemShut
  {NoStop}%
\bibitem [{\citenamefont {Ding}\ \emph {et~al.}(2021)\citenamefont {Ding},
  \citenamefont {Li}, \citenamefont {Tomiya}, \citenamefont {Wang},\ and\
  \citenamefont {Zhang}}]{Ding:2020hxw}%
  \BibitemOpen
  \bibfield  {author} {\bibinfo {author} {\bibfnamefont {H.~T.}\ \bibnamefont
  {Ding}}, \bibinfo {author} {\bibfnamefont {S.~T.}\ \bibnamefont {Li}},
  \bibinfo {author} {\bibfnamefont {A.}~\bibnamefont {Tomiya}}, \bibinfo
  {author} {\bibfnamefont {X.~D.}\ \bibnamefont {Wang}},\ and\ \bibinfo
  {author} {\bibfnamefont {Y.}~\bibnamefont {Zhang}},\ }\bibfield  {title}
  {\bibinfo {title} {{Chiral properties of (2+1)-flavor QCD in strong magnetic
  fields at zero temperature}},\ }\href
  {https://doi.org/10.1103/PhysRevD.104.014505} {\bibfield  {journal} {\bibinfo
   {journal} {Phys. Rev. D}\ }\textbf {\bibinfo {volume} {104}},\ \bibinfo
  {pages} {014505} (\bibinfo {year} {2021})},\ \Eprint
  {https://arxiv.org/abs/2008.00493} {arXiv:2008.00493 [hep-lat]} \BibitemShut
  {NoStop}%
\bibitem [{\citenamefont {Ding}\ \emph {et~al.}(2022)\citenamefont {Ding},
  \citenamefont {Li}, \citenamefont {Liu},\ and\ \citenamefont
  {Wang}}]{Ding:2022tqn}%
  \BibitemOpen
  \bibfield  {author} {\bibinfo {author} {\bibfnamefont {H.~T.}\ \bibnamefont
  {Ding}}, \bibinfo {author} {\bibfnamefont {S.~T.}\ \bibnamefont {Li}},
  \bibinfo {author} {\bibfnamefont {J.~H.}\ \bibnamefont {Liu}},\ and\ \bibinfo
  {author} {\bibfnamefont {X.~D.}\ \bibnamefont {Wang}},\ }\bibfield  {title}
  {\bibinfo {title} {{Chiral condensates and screening masses of neutral
  pseudoscalar mesons in thermomagnetic QCD medium}},\ }\href
  {https://doi.org/10.1103/PhysRevD.105.034514} {\bibfield  {journal} {\bibinfo
   {journal} {Phys. Rev. D}\ }\textbf {\bibinfo {volume} {105}},\ \bibinfo
  {pages} {034514} (\bibinfo {year} {2022})},\ \Eprint
  {https://arxiv.org/abs/2201.02349} {arXiv:2201.02349 [hep-lat]} \BibitemShut
  {NoStop}%
\bibitem [{\citenamefont {Inagaki}\ \emph {et~al.}(2004)\citenamefont
  {Inagaki}, \citenamefont {Kimura},\ and\ \citenamefont
  {Murata}}]{Inagaki:2003yi}%
  \BibitemOpen
  \bibfield  {author} {\bibinfo {author} {\bibfnamefont {T.}~\bibnamefont
  {Inagaki}}, \bibinfo {author} {\bibfnamefont {D.}~\bibnamefont {Kimura}},\
  and\ \bibinfo {author} {\bibfnamefont {T.}~\bibnamefont {Murata}},\
  }\bibfield  {title} {\bibinfo {title} {{Four fermion interaction model in a
  constant magnetic field at finite temperature and chemical potential}},\
  }\href {https://doi.org/10.1143/PTP.111.371} {\bibfield  {journal} {\bibinfo
  {journal} {Prog. Theor. Phys.}\ }\textbf {\bibinfo {volume} {111}},\ \bibinfo
  {pages} {371} (\bibinfo {year} {2004})},\ \Eprint
  {https://arxiv.org/abs/hep-ph/0312005} {arXiv:hep-ph/0312005} \BibitemShut
  {NoStop}%
\bibitem [{\citenamefont {Chao}\ \emph {et~al.}(2014)\citenamefont {Chao},
  \citenamefont {Yu},\ and\ \citenamefont {Huang}}]{Chao:2014wla}%
  \BibitemOpen
  \bibfield  {author} {\bibinfo {author} {\bibfnamefont {J.}~\bibnamefont
  {Chao}}, \bibinfo {author} {\bibfnamefont {L.}~\bibnamefont {Yu}},\ and\
  \bibinfo {author} {\bibfnamefont {M.}~\bibnamefont {Huang}},\ }\bibfield
  {title} {\bibinfo {title} {{Zeta function regularization of the photon
  polarization tensor for a magnetized vacuum}},\ }\href
  {https://doi.org/10.1103/PhysRevD.90.045033} {\bibfield  {journal} {\bibinfo
  {journal} {Phys. Rev. D}\ }\textbf {\bibinfo {volume} {90}},\ \bibinfo
  {pages} {045033} (\bibinfo {year} {2014})},\ \bibinfo {note} {[Erratum:
  Phys.Rev.D 91, 029903 (2015)]},\ \Eprint {https://arxiv.org/abs/1403.0442}
  {arXiv:1403.0442 [hep-th]} \BibitemShut {NoStop}%
\bibitem [{\citenamefont {Yu}\ \emph {et~al.}(2015)\citenamefont {Yu},
  \citenamefont {Van~Doorsselaere},\ and\ \citenamefont {Huang}}]{Yu:2014xoa}%
  \BibitemOpen
  \bibfield  {author} {\bibinfo {author} {\bibfnamefont {L.}~\bibnamefont
  {Yu}}, \bibinfo {author} {\bibfnamefont {J.}~\bibnamefont
  {Van~Doorsselaere}},\ and\ \bibinfo {author} {\bibfnamefont {M.}~\bibnamefont
  {Huang}},\ }\bibfield  {title} {\bibinfo {title} {{Inverse Magnetic Catalysis
  in the three-flavor NJL model with axial-vector interaction}},\ }\href
  {https://doi.org/10.1103/PhysRevD.91.074011} {\bibfield  {journal} {\bibinfo
  {journal} {Phys. Rev. D}\ }\textbf {\bibinfo {volume} {91}},\ \bibinfo
  {pages} {074011} (\bibinfo {year} {2015})},\ \Eprint
  {https://arxiv.org/abs/1411.7552} {arXiv:1411.7552 [hep-ph]} \BibitemShut
  {NoStop}%
\bibitem [{\citenamefont {Coppola}\ \emph {et~al.}(2018)\citenamefont
  {Coppola}, \citenamefont {G\'omez~Dumm},\ and\ \citenamefont
  {Scoccola}}]{Coppola:2018vkw}%
  \BibitemOpen
  \bibfield  {author} {\bibinfo {author} {\bibfnamefont {M.}~\bibnamefont
  {Coppola}}, \bibinfo {author} {\bibfnamefont {D.}~\bibnamefont
  {G\'omez~Dumm}},\ and\ \bibinfo {author} {\bibfnamefont {N.~N.}\ \bibnamefont
  {Scoccola}},\ }\bibfield  {title} {\bibinfo {title} {{Charged pion masses
  under strong magnetic fields in the NJL model}},\ }\href
  {https://doi.org/10.1016/j.physletb.2018.04.043} {\bibfield  {journal}
  {\bibinfo  {journal} {Phys. Lett. B}\ }\textbf {\bibinfo {volume} {782}},\
  \bibinfo {pages} {155} (\bibinfo {year} {2018})},\ \Eprint
  {https://arxiv.org/abs/1802.08041} {arXiv:1802.08041 [hep-ph]} \BibitemShut
  {NoStop}%
\bibitem [{\citenamefont {Coppola}\ \emph {et~al.}(2019)\citenamefont
  {Coppola}, \citenamefont {Gomez~Dumm}, \citenamefont {Noguera},\ and\
  \citenamefont {Scoccola}}]{Coppola:2019uyr}%
  \BibitemOpen
  \bibfield  {author} {\bibinfo {author} {\bibfnamefont {M.}~\bibnamefont
  {Coppola}}, \bibinfo {author} {\bibfnamefont {D.}~\bibnamefont {Gomez~Dumm}},
  \bibinfo {author} {\bibfnamefont {S.}~\bibnamefont {Noguera}},\ and\ \bibinfo
  {author} {\bibfnamefont {N.~N.}\ \bibnamefont {Scoccola}},\ }\bibfield
  {title} {\bibinfo {title} {{Neutral and charged pion properties under strong
  magnetic fields in the NJL model}},\ }\href
  {https://doi.org/10.1103/PhysRevD.100.054014} {\bibfield  {journal} {\bibinfo
   {journal} {Phys. Rev. D}\ }\textbf {\bibinfo {volume} {100}},\ \bibinfo
  {pages} {054014} (\bibinfo {year} {2019})},\ \Eprint
  {https://arxiv.org/abs/1907.05840} {arXiv:1907.05840 [hep-ph]} \BibitemShut
  {NoStop}%
\bibitem [{\citenamefont {Fayazbakhsh}\ and\ \citenamefont
  {Sadooghi}(2014)}]{Fayazbakhsh:2014mca}%
  \BibitemOpen
  \bibfield  {author} {\bibinfo {author} {\bibfnamefont {S.}~\bibnamefont
  {Fayazbakhsh}}\ and\ \bibinfo {author} {\bibfnamefont {N.}~\bibnamefont
  {Sadooghi}},\ }\bibfield  {title} {\bibinfo {title} {{Anomalous magnetic
  moment of hot quarks, inverse magnetic catalysis, and reentrance of the
  chiral symmetry broken phase}},\ }\href
  {https://doi.org/10.1103/PhysRevD.90.105030} {\bibfield  {journal} {\bibinfo
  {journal} {Phys. Rev. D}\ }\textbf {\bibinfo {volume} {90}},\ \bibinfo
  {pages} {105030} (\bibinfo {year} {2014})},\ \Eprint
  {https://arxiv.org/abs/1408.5457} {arXiv:1408.5457 [hep-ph]} \BibitemShut
  {NoStop}%
\bibitem [{\citenamefont {Chaudhuri}\ \emph {et~al.}(2019)\citenamefont
  {Chaudhuri}, \citenamefont {Ghosh}, \citenamefont {Sarkar},\ and\
  \citenamefont {Roy}}]{Chaudhuri:2019lbw}%
  \BibitemOpen
  \bibfield  {author} {\bibinfo {author} {\bibfnamefont {N.}~\bibnamefont
  {Chaudhuri}}, \bibinfo {author} {\bibfnamefont {S.}~\bibnamefont {Ghosh}},
  \bibinfo {author} {\bibfnamefont {S.}~\bibnamefont {Sarkar}},\ and\ \bibinfo
  {author} {\bibfnamefont {P.}~\bibnamefont {Roy}},\ }\bibfield  {title}
  {\bibinfo {title} {{Effect of the anomalous magnetic moment of quarks on the
  phase structure and mesonic properties in the NJL model}},\ }\href
  {https://doi.org/10.1103/PhysRevD.99.116025} {\bibfield  {journal} {\bibinfo
  {journal} {Phys. Rev. D}\ }\textbf {\bibinfo {volume} {99}},\ \bibinfo
  {pages} {116025} (\bibinfo {year} {2019})},\ \Eprint
  {https://arxiv.org/abs/1907.03990} {arXiv:1907.03990 [nucl-th]} \BibitemShut
  {NoStop}%
\bibitem [{\citenamefont {Ayala}\ \emph {et~al.}(2018)\citenamefont {Ayala},
  \citenamefont {Farias}, \citenamefont {Hern\'andez-Ortiz}, \citenamefont
  {Hern\'andez}, \citenamefont {Paret},\ and\ \citenamefont
  {Zamora}}]{Ayala:2018zat}%
  \BibitemOpen
  \bibfield  {author} {\bibinfo {author} {\bibfnamefont {A.}~\bibnamefont
  {Ayala}}, \bibinfo {author} {\bibfnamefont {R.~L.~S.}\ \bibnamefont
  {Farias}}, \bibinfo {author} {\bibfnamefont {S.}~\bibnamefont
  {Hern\'andez-Ortiz}}, \bibinfo {author} {\bibfnamefont {L.~A.}\ \bibnamefont
  {Hern\'andez}}, \bibinfo {author} {\bibfnamefont {D.~M.}\ \bibnamefont
  {Paret}},\ and\ \bibinfo {author} {\bibfnamefont {R.}~\bibnamefont
  {Zamora}},\ }\bibfield  {title} {\bibinfo {title} {{Magnetic field-dependence
  of the neutral pion mass in the linear sigma model coupled to quarks: The
  weak field case}},\ }\href {https://doi.org/10.1103/PhysRevD.98.114008}
  {\bibfield  {journal} {\bibinfo  {journal} {Phys. Rev. D}\ }\textbf {\bibinfo
  {volume} {98}},\ \bibinfo {pages} {114008} (\bibinfo {year} {2018})},\
  \Eprint {https://arxiv.org/abs/1809.08312} {arXiv:1809.08312 [hep-ph]}
  \BibitemShut {NoStop}%
\bibitem [{\citenamefont {Kamikado}\ and\ \citenamefont
  {Kanazawa}(2014)}]{Kamikado:2013pya}%
  \BibitemOpen
  \bibfield  {author} {\bibinfo {author} {\bibfnamefont {K.}~\bibnamefont
  {Kamikado}}\ and\ \bibinfo {author} {\bibfnamefont {T.}~\bibnamefont
  {Kanazawa}},\ }\bibfield  {title} {\bibinfo {title} {{Chiral dynamics in a
  magnetic field from the functional renormalization group}},\ }\href
  {https://doi.org/10.1007/JHEP03(2014)009} {\bibfield  {journal} {\bibinfo
  {journal} {JHEP}\ }\textbf {\bibinfo {volume} {03}},\ \bibinfo {pages}
  {009}},\ \Eprint {https://arxiv.org/abs/1312.3124} {arXiv:1312.3124 [hep-ph]}
  \BibitemShut {NoStop}%
\bibitem [{\citenamefont {Kamikado}\ and\ \citenamefont
  {Kanazawa}(2015)}]{Kamikado:2014bua}%
  \BibitemOpen
  \bibfield  {author} {\bibinfo {author} {\bibfnamefont {K.}~\bibnamefont
  {Kamikado}}\ and\ \bibinfo {author} {\bibfnamefont {T.}~\bibnamefont
  {Kanazawa}},\ }\bibfield  {title} {\bibinfo {title} {{Magnetic susceptibility
  of a strongly interacting thermal medium with 2$+$1 quark flavors}},\ }\href
  {https://doi.org/10.1007/JHEP01(2015)129} {\bibfield  {journal} {\bibinfo
  {journal} {JHEP}\ }\textbf {\bibinfo {volume} {01}},\ \bibinfo {pages}
  {129}},\ \Eprint {https://arxiv.org/abs/1410.6253} {arXiv:1410.6253 [hep-ph]}
  \BibitemShut {NoStop}%
\bibitem [{\citenamefont {Mamo}(2015)}]{Mamo:2015dea}%
  \BibitemOpen
  \bibfield  {author} {\bibinfo {author} {\bibfnamefont {K.~A.}\ \bibnamefont
  {Mamo}},\ }\bibfield  {title} {\bibinfo {title} {{Inverse magnetic catalysis
  in holographic models of QCD}},\ }\href
  {https://doi.org/10.1007/JHEP05(2015)121} {\bibfield  {journal} {\bibinfo
  {journal} {JHEP}\ }\textbf {\bibinfo {volume} {05}},\ \bibinfo {pages}
  {121}},\ \Eprint {https://arxiv.org/abs/1501.03262} {arXiv:1501.03262
  [hep-th]} \BibitemShut {NoStop}%
\bibitem [{\citenamefont {Li}\ \emph {et~al.}(2017)\citenamefont {Li},
  \citenamefont {Huang}, \citenamefont {Yang},\ and\ \citenamefont
  {Yuan}}]{Li:2016gfn}%
  \BibitemOpen
  \bibfield  {author} {\bibinfo {author} {\bibfnamefont {D.}~\bibnamefont
  {Li}}, \bibinfo {author} {\bibfnamefont {M.}~\bibnamefont {Huang}}, \bibinfo
  {author} {\bibfnamefont {Y.}~\bibnamefont {Yang}},\ and\ \bibinfo {author}
  {\bibfnamefont {P.-H.}\ \bibnamefont {Yuan}},\ }\bibfield  {title} {\bibinfo
  {title} {{Inverse Magnetic Catalysis in the Soft-Wall Model of AdS/QCD}},\
  }\href {https://doi.org/10.1007/JHEP02(2017)030} {\bibfield  {journal}
  {\bibinfo  {journal} {JHEP}\ }\textbf {\bibinfo {volume} {02}},\ \bibinfo
  {pages} {030}},\ \Eprint {https://arxiv.org/abs/1610.04618} {arXiv:1610.04618
  [hep-th]} \BibitemShut {NoStop}%
\bibitem [{\citenamefont {Fukushima}\ and\ \citenamefont
  {Pawlowski}(2012)}]{Fukushima:2012xw}%
  \BibitemOpen
  \bibfield  {author} {\bibinfo {author} {\bibfnamefont {K.}~\bibnamefont
  {Fukushima}}\ and\ \bibinfo {author} {\bibfnamefont {J.~M.}\ \bibnamefont
  {Pawlowski}},\ }\bibfield  {title} {\bibinfo {title} {{Magnetic catalysis in
  hot and dense quark matter and quantum fluctuations}},\ }\href
  {https://doi.org/10.1103/PhysRevD.86.076013} {\bibfield  {journal} {\bibinfo
  {journal} {Phys. Rev. D}\ }\textbf {\bibinfo {volume} {86}},\ \bibinfo
  {pages} {076013} (\bibinfo {year} {2012})},\ \Eprint
  {https://arxiv.org/abs/1203.4330} {arXiv:1203.4330 [hep-ph]} \BibitemShut
  {NoStop}%
\bibitem [{\citenamefont {Braun}\ \emph
  {et~al.}(2016{\natexlab{a}})\citenamefont {Braun}, \citenamefont {Mian},\
  and\ \citenamefont {Rechenberger}}]{Braun:2014fua}%
  \BibitemOpen
  \bibfield  {author} {\bibinfo {author} {\bibfnamefont {J.}~\bibnamefont
  {Braun}}, \bibinfo {author} {\bibfnamefont {W.~A.}\ \bibnamefont {Mian}},\
  and\ \bibinfo {author} {\bibfnamefont {S.}~\bibnamefont {Rechenberger}},\
  }\bibfield  {title} {\bibinfo {title} {{Delayed Magnetic Catalysis}},\ }\href
  {https://doi.org/10.1016/j.physletb.2016.02.017} {\bibfield  {journal}
  {\bibinfo  {journal} {Phys. Lett. B}\ }\textbf {\bibinfo {volume} {755}},\
  \bibinfo {pages} {265} (\bibinfo {year} {2016}{\natexlab{a}})},\ \Eprint
  {https://arxiv.org/abs/1412.6025} {arXiv:1412.6025 [hep-ph]} \BibitemShut
  {NoStop}%
\bibitem [{\citenamefont {Mueller}\ and\ \citenamefont
  {Pawlowski}(2015)}]{Mueller:2015fka}%
  \BibitemOpen
  \bibfield  {author} {\bibinfo {author} {\bibfnamefont {N.}~\bibnamefont
  {Mueller}}\ and\ \bibinfo {author} {\bibfnamefont {J.~M.}\ \bibnamefont
  {Pawlowski}},\ }\bibfield  {title} {\bibinfo {title} {{Magnetic catalysis and
  inverse magnetic catalysis in QCD}},\ }\href
  {https://doi.org/10.1103/PhysRevD.91.116010} {\bibfield  {journal} {\bibinfo
  {journal} {Phys. Rev. D}\ }\textbf {\bibinfo {volume} {91}},\ \bibinfo
  {pages} {116010} (\bibinfo {year} {2015})},\ \Eprint
  {https://arxiv.org/abs/1502.08011} {arXiv:1502.08011 [hep-ph]} \BibitemShut
  {NoStop}%
\bibitem [{\citenamefont {Fu}\ and\ \citenamefont {Liu}(2017)}]{Fu:2017vvg}%
  \BibitemOpen
  \bibfield  {author} {\bibinfo {author} {\bibfnamefont {W.-j.}\ \bibnamefont
  {Fu}}\ and\ \bibinfo {author} {\bibfnamefont {Y.-x.}\ \bibnamefont {Liu}},\
  }\bibfield  {title} {\bibinfo {title} {{Four-fermion interactions and the
  chiral symmetry breaking in an external magnetic field}},\ }\href
  {https://doi.org/10.1103/PhysRevD.96.074019} {\bibfield  {journal} {\bibinfo
  {journal} {Phys. Rev. D}\ }\textbf {\bibinfo {volume} {96}},\ \bibinfo
  {pages} {074019} (\bibinfo {year} {2017})},\ \Eprint
  {https://arxiv.org/abs/1705.09841} {arXiv:1705.09841 [hep-ph]} \BibitemShut
  {NoStop}%
\bibitem [{\citenamefont {Li}\ \emph {et~al.}(2019)\citenamefont {Li},
  \citenamefont {Fu},\ and\ \citenamefont {Liu}}]{Li:2019nzj}%
  \BibitemOpen
  \bibfield  {author} {\bibinfo {author} {\bibfnamefont {X.}~\bibnamefont
  {Li}}, \bibinfo {author} {\bibfnamefont {W.-J.}\ \bibnamefont {Fu}},\ and\
  \bibinfo {author} {\bibfnamefont {Y.-X.}\ \bibnamefont {Liu}},\ }\bibfield
  {title} {\bibinfo {title} {{Thermodynamics of 2+1 Flavor Polyakov-Loop
  Quark-Meson Model under External Magnetic Field}},\ }\href
  {https://doi.org/10.1103/PhysRevD.99.074029} {\bibfield  {journal} {\bibinfo
  {journal} {Phys. Rev. D}\ }\textbf {\bibinfo {volume} {99}},\ \bibinfo
  {pages} {074029} (\bibinfo {year} {2019})},\ \Eprint
  {https://arxiv.org/abs/1902.03866} {arXiv:1902.03866 [hep-ph]} \BibitemShut
  {NoStop}%
\bibitem [{\citenamefont {Shovkovy}(2013)}]{Shovkovy:2012zn}%
  \BibitemOpen
  \bibfield  {author} {\bibinfo {author} {\bibfnamefont {I.~A.}\ \bibnamefont
  {Shovkovy}},\ }\bibfield  {title} {\bibinfo {title} {{Magnetic Catalysis: A
  Review}},\ }\href {https://doi.org/10.1007/978-3-642-37305-3_2} {\bibfield
  {journal} {\bibinfo  {journal} {Lect. Notes Phys.}\ }\textbf {\bibinfo
  {volume} {871}},\ \bibinfo {pages} {13} (\bibinfo {year} {2013})},\ \Eprint
  {https://arxiv.org/abs/1207.5081} {arXiv:1207.5081 [hep-ph]} \BibitemShut
  {NoStop}%
\bibitem [{\citenamefont {Andersen}\ \emph {et~al.}(2016)\citenamefont
  {Andersen}, \citenamefont {Naylor},\ and\ \citenamefont
  {Tranberg}}]{Andersen:2014xxa}%
  \BibitemOpen
  \bibfield  {author} {\bibinfo {author} {\bibfnamefont {J.~O.}\ \bibnamefont
  {Andersen}}, \bibinfo {author} {\bibfnamefont {W.~R.}\ \bibnamefont
  {Naylor}},\ and\ \bibinfo {author} {\bibfnamefont {A.}~\bibnamefont
  {Tranberg}},\ }\bibfield  {title} {\bibinfo {title} {{Phase diagram of QCD in
  a magnetic field: A review}},\ }\href
  {https://doi.org/10.1103/RevModPhys.88.025001} {\bibfield  {journal}
  {\bibinfo  {journal} {Rev. Mod. Phys.}\ }\textbf {\bibinfo {volume} {88}},\
  \bibinfo {pages} {025001} (\bibinfo {year} {2016})},\ \Eprint
  {https://arxiv.org/abs/1411.7176} {arXiv:1411.7176 [hep-ph]} \BibitemShut
  {NoStop}%
\bibitem [{\citenamefont {Miransky}\ and\ \citenamefont
  {Shovkovy}(2015)}]{Miransky:2015ava}%
  \BibitemOpen
  \bibfield  {author} {\bibinfo {author} {\bibfnamefont {V.~A.}\ \bibnamefont
  {Miransky}}\ and\ \bibinfo {author} {\bibfnamefont {I.~A.}\ \bibnamefont
  {Shovkovy}},\ }\bibfield  {title} {\bibinfo {title} {{Quantum field theory in
  a magnetic field: From quantum chromodynamics to graphene and Dirac
  semimetals}},\ }\href {https://doi.org/10.1016/j.physrep.2015.02.003}
  {\bibfield  {journal} {\bibinfo  {journal} {Phys. Rept.}\ }\textbf {\bibinfo
  {volume} {576}},\ \bibinfo {pages} {1} (\bibinfo {year} {2015})},\ \Eprint
  {https://arxiv.org/abs/1503.00732} {arXiv:1503.00732 [hep-ph]} \BibitemShut
  {NoStop}%
\bibitem [{\citenamefont {Hattori}\ \emph {et~al.}(2023)\citenamefont
  {Hattori}, \citenamefont {Itakura},\ and\ \citenamefont
  {Ozaki}}]{Hattori:2023egw}%
  \BibitemOpen
  \bibfield  {author} {\bibinfo {author} {\bibfnamefont {K.}~\bibnamefont
  {Hattori}}, \bibinfo {author} {\bibfnamefont {K.}~\bibnamefont {Itakura}},\
  and\ \bibinfo {author} {\bibfnamefont {S.}~\bibnamefont {Ozaki}},\ }\bibfield
   {title} {\bibinfo {title} {{Strong-Field Physics in QED and QCD: From
  Fundamentals to Applications}},\ }\href@noop {} {\  (\bibinfo {year}
  {2023})},\ \Eprint {https://arxiv.org/abs/2305.03865} {arXiv:2305.03865
  [hep-ph]} \BibitemShut {NoStop}%
\bibitem [{\citenamefont {Fukushima}\ and\ \citenamefont
  {Hidaka}(2013)}]{Fukushima:2012kc}%
  \BibitemOpen
  \bibfield  {author} {\bibinfo {author} {\bibfnamefont {K.}~\bibnamefont
  {Fukushima}}\ and\ \bibinfo {author} {\bibfnamefont {Y.}~\bibnamefont
  {Hidaka}},\ }\bibfield  {title} {\bibinfo {title} {{Magnetic Catalysis Versus
  Magnetic Inhibition}},\ }\href
  {https://doi.org/10.1103/PhysRevLett.110.031601} {\bibfield  {journal}
  {\bibinfo  {journal} {Phys. Rev. Lett.}\ }\textbf {\bibinfo {volume} {110}},\
  \bibinfo {pages} {031601} (\bibinfo {year} {2013})},\ \Eprint
  {https://arxiv.org/abs/1209.1319} {arXiv:1209.1319 [hep-ph]} \BibitemShut
  {NoStop}%
\bibitem [{\citenamefont {Mao}(2016)}]{Mao:2016fha}%
  \BibitemOpen
  \bibfield  {author} {\bibinfo {author} {\bibfnamefont {S.}~\bibnamefont
  {Mao}},\ }\bibfield  {title} {\bibinfo {title} {{Inverse magnetic catalysis
  in Nambu\textendash{}Jona-Lasinio model beyond mean field}},\ }\href
  {https://doi.org/10.1016/j.physletb.2016.05.018} {\bibfield  {journal}
  {\bibinfo  {journal} {Phys. Lett. B}\ }\textbf {\bibinfo {volume} {758}},\
  \bibinfo {pages} {195} (\bibinfo {year} {2016})},\ \Eprint
  {https://arxiv.org/abs/1602.06503} {arXiv:1602.06503 [hep-ph]} \BibitemShut
  {NoStop}%
\bibitem [{\citenamefont {Luschevskaya}\ \emph {et~al.}(2017)\citenamefont
  {Luschevskaya}, \citenamefont {Solovjeva},\ and\ \citenamefont
  {Teryaev}}]{Luschevskaya:2016epp}%
  \BibitemOpen
  \bibfield  {author} {\bibinfo {author} {\bibfnamefont {E.~V.}\ \bibnamefont
  {Luschevskaya}}, \bibinfo {author} {\bibfnamefont {O.~E.}\ \bibnamefont
  {Solovjeva}},\ and\ \bibinfo {author} {\bibfnamefont {O.~V.}\ \bibnamefont
  {Teryaev}},\ }\bibfield  {title} {\bibinfo {title} {{Determination of the
  properties of vector mesons in external magnetic field by Quenched $SU(3)$
  Lattice QCD}},\ }\href {https://doi.org/10.1007/JHEP09(2017)142} {\bibfield
  {journal} {\bibinfo  {journal} {JHEP}\ }\textbf {\bibinfo {volume} {09}},\
  \bibinfo {pages} {142}},\ \Eprint {https://arxiv.org/abs/1608.03472}
  {arXiv:1608.03472 [hep-lat]} \BibitemShut {NoStop}%
\bibitem [{\citenamefont {Wang}\ and\ \citenamefont
  {Zhuang}(2018)}]{Wang:2017vtn}%
  \BibitemOpen
  \bibfield  {author} {\bibinfo {author} {\bibfnamefont {Z.}~\bibnamefont
  {Wang}}\ and\ \bibinfo {author} {\bibfnamefont {P.}~\bibnamefont {Zhuang}},\
  }\bibfield  {title} {\bibinfo {title} {{Meson properties in magnetized quark
  matter}},\ }\href {https://doi.org/10.1103/PhysRevD.97.034026} {\bibfield
  {journal} {\bibinfo  {journal} {Phys. Rev. D}\ }\textbf {\bibinfo {volume}
  {97}},\ \bibinfo {pages} {034026} (\bibinfo {year} {2018})},\ \Eprint
  {https://arxiv.org/abs/1712.00554} {arXiv:1712.00554 [hep-ph]} \BibitemShut
  {NoStop}%
\bibitem [{\citenamefont {Liu}\ \emph {et~al.}(2018)\citenamefont {Liu},
  \citenamefont {Wang}, \citenamefont {Yu},\ and\ \citenamefont
  {Huang}}]{Liu:2018zag}%
  \BibitemOpen
  \bibfield  {author} {\bibinfo {author} {\bibfnamefont {H.}~\bibnamefont
  {Liu}}, \bibinfo {author} {\bibfnamefont {X.}~\bibnamefont {Wang}}, \bibinfo
  {author} {\bibfnamefont {L.}~\bibnamefont {Yu}},\ and\ \bibinfo {author}
  {\bibfnamefont {M.}~\bibnamefont {Huang}},\ }\bibfield  {title} {\bibinfo
  {title} {{Neutral and charged scalar mesons, pseudoscalar mesons, and
  diquarks in magnetic fields}},\ }\href
  {https://doi.org/10.1103/PhysRevD.97.076008} {\bibfield  {journal} {\bibinfo
  {journal} {Phys. Rev. D}\ }\textbf {\bibinfo {volume} {97}},\ \bibinfo
  {pages} {076008} (\bibinfo {year} {2018})},\ \Eprint
  {https://arxiv.org/abs/1801.02174} {arXiv:1801.02174 [hep-ph]} \BibitemShut
  {NoStop}%
\bibitem [{\citenamefont {Ayala}\ \emph {et~al.}(2021)\citenamefont {Ayala},
  \citenamefont {Hern\'andez}, \citenamefont {Hern\'andez}, \citenamefont
  {Farias},\ and\ \citenamefont {Zamora}}]{Ayala:2020dxs}%
  \BibitemOpen
  \bibfield  {author} {\bibinfo {author} {\bibfnamefont {A.}~\bibnamefont
  {Ayala}}, \bibinfo {author} {\bibfnamefont {J.~L.}\ \bibnamefont
  {Hern\'andez}}, \bibinfo {author} {\bibfnamefont {L.~A.}\ \bibnamefont
  {Hern\'andez}}, \bibinfo {author} {\bibfnamefont {R.~L.~S.}\ \bibnamefont
  {Farias}},\ and\ \bibinfo {author} {\bibfnamefont {R.}~\bibnamefont
  {Zamora}},\ }\bibfield  {title} {\bibinfo {title} {{Magnetic field dependence
  of the neutral pion mass in the linear sigma model with quarks: The strong
  field case}},\ }\href {https://doi.org/10.1103/PhysRevD.103.054038}
  {\bibfield  {journal} {\bibinfo  {journal} {Phys. Rev. D}\ }\textbf {\bibinfo
  {volume} {103}},\ \bibinfo {pages} {054038} (\bibinfo {year} {2021})},\
  \Eprint {https://arxiv.org/abs/2011.03673} {arXiv:2011.03673 [hep-ph]}
  \BibitemShut {NoStop}%
\bibitem [{\citenamefont {Hidaka}\ and\ \citenamefont
  {Yamamoto}(2013)}]{Hidaka:2012mz}%
  \BibitemOpen
  \bibfield  {author} {\bibinfo {author} {\bibfnamefont {Y.}~\bibnamefont
  {Hidaka}}\ and\ \bibinfo {author} {\bibfnamefont {A.}~\bibnamefont
  {Yamamoto}},\ }\bibfield  {title} {\bibinfo {title} {{Charged vector mesons
  in a strong magnetic field}},\ }\href
  {https://doi.org/10.1103/PhysRevD.87.094502} {\bibfield  {journal} {\bibinfo
  {journal} {Phys. Rev. D}\ }\textbf {\bibinfo {volume} {87}},\ \bibinfo
  {pages} {094502} (\bibinfo {year} {2013})},\ \Eprint
  {https://arxiv.org/abs/1209.0007} {arXiv:1209.0007 [hep-ph]} \BibitemShut
  {NoStop}%
\bibitem [{\citenamefont {Li}\ \emph {et~al.}(2021)\citenamefont {Li},
  \citenamefont {Cao},\ and\ \citenamefont {He}}]{Li:2020hlp}%
  \BibitemOpen
  \bibfield  {author} {\bibinfo {author} {\bibfnamefont {J.}~\bibnamefont
  {Li}}, \bibinfo {author} {\bibfnamefont {G.}~\bibnamefont {Cao}},\ and\
  \bibinfo {author} {\bibfnamefont {L.}~\bibnamefont {He}},\ }\bibfield
  {title} {\bibinfo {title} {{Gauge independence of pion masses in a magnetic
  field within the Nambu\textendash{}Jona-Lasinio model}},\ }\href
  {https://doi.org/10.1103/PhysRevD.104.074026} {\bibfield  {journal} {\bibinfo
   {journal} {Phys. Rev. D}\ }\textbf {\bibinfo {volume} {104}},\ \bibinfo
  {pages} {074026} (\bibinfo {year} {2021})},\ \Eprint
  {https://arxiv.org/abs/2009.04697} {arXiv:2009.04697 [nucl-th]} \BibitemShut
  {NoStop}%
\bibitem [{\citenamefont {Carlomagno}\ \emph {et~al.}(2022)\citenamefont
  {Carlomagno}, \citenamefont {Gomez~Dumm}, \citenamefont {Villafa\~ne},
  \citenamefont {Noguera},\ and\ \citenamefont
  {Scoccola}}]{Carlomagno:2022arc}%
  \BibitemOpen
  \bibfield  {author} {\bibinfo {author} {\bibfnamefont {J.~P.}\ \bibnamefont
  {Carlomagno}}, \bibinfo {author} {\bibfnamefont {D.}~\bibnamefont
  {Gomez~Dumm}}, \bibinfo {author} {\bibfnamefont {M.~F.~I.}\ \bibnamefont
  {Villafa\~ne}}, \bibinfo {author} {\bibfnamefont {S.}~\bibnamefont
  {Noguera}},\ and\ \bibinfo {author} {\bibfnamefont {N.~N.}\ \bibnamefont
  {Scoccola}},\ }\bibfield  {title} {\bibinfo {title} {{Charged pseudoscalar
  and vector meson masses in strong magnetic fields in an extended NJL
  model}},\ }\href {https://doi.org/10.1103/PhysRevD.106.094035} {\bibfield
  {journal} {\bibinfo  {journal} {Phys. Rev. D}\ }\textbf {\bibinfo {volume}
  {106}},\ \bibinfo {pages} {094035} (\bibinfo {year} {2022})},\ \Eprint
  {https://arxiv.org/abs/2209.10679} {arXiv:2209.10679 [hep-ph]} \BibitemShut
  {NoStop}%
\bibitem [{\citenamefont {Endr\H{o}di}\ and\ \citenamefont
  {Mark\'o}(2019)}]{Endrodi:2019whh}%
  \BibitemOpen
  \bibfield  {author} {\bibinfo {author} {\bibfnamefont {G.}~\bibnamefont
  {Endr\H{o}di}}\ and\ \bibinfo {author} {\bibfnamefont {G.}~\bibnamefont
  {Mark\'o}},\ }\bibfield  {title} {\bibinfo {title} {{Magnetized baryons and
  the QCD phase diagram: NJL model meets the lattice}},\ }\href
  {https://doi.org/10.1007/JHEP08(2019)036} {\bibfield  {journal} {\bibinfo
  {journal} {JHEP}\ }\textbf {\bibinfo {volume} {08}},\ \bibinfo {pages}
  {036}},\ \Eprint {https://arxiv.org/abs/1905.02103} {arXiv:1905.02103
  [hep-lat]} \BibitemShut {NoStop}%
\bibitem [{\citenamefont {Xu}\ \emph {et~al.}(2021)\citenamefont {Xu},
  \citenamefont {Chao},\ and\ \citenamefont {Huang}}]{Xu:2020yag}%
  \BibitemOpen
  \bibfield  {author} {\bibinfo {author} {\bibfnamefont {K.}~\bibnamefont
  {Xu}}, \bibinfo {author} {\bibfnamefont {J.}~\bibnamefont {Chao}},\ and\
  \bibinfo {author} {\bibfnamefont {M.}~\bibnamefont {Huang}},\ }\bibfield
  {title} {\bibinfo {title} {{Effect of the anomalous magnetic moment of quarks
  on magnetized QCD matter and meson spectra}},\ }\href
  {https://doi.org/10.1103/PhysRevD.103.076015} {\bibfield  {journal} {\bibinfo
   {journal} {Phys. Rev. D}\ }\textbf {\bibinfo {volume} {103}},\ \bibinfo
  {pages} {076015} (\bibinfo {year} {2021})},\ \Eprint
  {https://arxiv.org/abs/2007.13122} {arXiv:2007.13122 [hep-ph]} \BibitemShut
  {NoStop}%
\bibitem [{\citenamefont {Lin}\ \emph {et~al.}(2022)\citenamefont {Lin},
  \citenamefont {Xu},\ and\ \citenamefont {Huang}}]{Lin:2022ied}%
  \BibitemOpen
  \bibfield  {author} {\bibinfo {author} {\bibfnamefont {F.}~\bibnamefont
  {Lin}}, \bibinfo {author} {\bibfnamefont {K.}~\bibnamefont {Xu}},\ and\
  \bibinfo {author} {\bibfnamefont {M.}~\bibnamefont {Huang}},\ }\bibfield
  {title} {\bibinfo {title} {{Magnetism of QCD matter and the pion mass from
  tensor-type spin polarization and the anomalous magnetic moment of quarks}},\
  }\href {https://doi.org/10.1103/PhysRevD.106.016005} {\bibfield  {journal}
  {\bibinfo  {journal} {Phys. Rev. D}\ }\textbf {\bibinfo {volume} {106}},\
  \bibinfo {pages} {016005} (\bibinfo {year} {2022})},\ \Eprint
  {https://arxiv.org/abs/2202.03226} {arXiv:2202.03226 [hep-ph]} \BibitemShut
  {NoStop}%
\bibitem [{\citenamefont {Xing}\ \emph {et~al.}(2022)\citenamefont {Xing},
  \citenamefont {Chao}, \citenamefont {Chang},\ and\ \citenamefont
  {Liu}}]{Xing:2021kbw}%
  \BibitemOpen
  \bibfield  {author} {\bibinfo {author} {\bibfnamefont {Z.}~\bibnamefont
  {Xing}}, \bibinfo {author} {\bibfnamefont {J.}~\bibnamefont {Chao}}, \bibinfo
  {author} {\bibfnamefont {L.}~\bibnamefont {Chang}},\ and\ \bibinfo {author}
  {\bibfnamefont {Y.-x.}\ \bibnamefont {Liu}},\ }\bibfield  {title} {\bibinfo
  {title} {{Exposing the effect of the p-wave component in the pion triplet
  under a strong magnetic field}},\ }\href
  {https://doi.org/10.1103/PhysRevD.105.114003} {\bibfield  {journal} {\bibinfo
   {journal} {Phys. Rev. D}\ }\textbf {\bibinfo {volume} {105}},\ \bibinfo
  {pages} {114003} (\bibinfo {year} {2022})},\ \Eprint
  {https://arxiv.org/abs/2110.01245} {arXiv:2110.01245 [hep-ph]} \BibitemShut
  {NoStop}%
\bibitem [{\citenamefont {Kojo}(2021)}]{Kojo:2021gvm}%
  \BibitemOpen
  \bibfield  {author} {\bibinfo {author} {\bibfnamefont {T.}~\bibnamefont
  {Kojo}},\ }\bibfield  {title} {\bibinfo {title} {{Neutral and charged mesons
  in magnetic fields: A resonance gas in a non-relativistic quark model}},\
  }\href {https://doi.org/10.1140/epja/s10050-021-00629-y} {\bibfield
  {journal} {\bibinfo  {journal} {Eur. Phys. J. A}\ }\textbf {\bibinfo {volume}
  {57}},\ \bibinfo {pages} {317} (\bibinfo {year} {2021})},\ \Eprint
  {https://arxiv.org/abs/2104.00376} {arXiv:2104.00376 [hep-ph]} \BibitemShut
  {NoStop}%
\bibitem [{\citenamefont {Sheng}\ \emph {et~al.}(2021)\citenamefont {Sheng},
  \citenamefont {Wang}, \citenamefont {Wang},\ and\ \citenamefont
  {Yu}}]{Sheng:2020hge}%
  \BibitemOpen
  \bibfield  {author} {\bibinfo {author} {\bibfnamefont {B.}~\bibnamefont
  {Sheng}}, \bibinfo {author} {\bibfnamefont {Y.}~\bibnamefont {Wang}},
  \bibinfo {author} {\bibfnamefont {X.}~\bibnamefont {Wang}},\ and\ \bibinfo
  {author} {\bibfnamefont {L.}~\bibnamefont {Yu}},\ }\bibfield  {title}
  {\bibinfo {title} {{Pole and screening masses of neutral pions in a hot and
  magnetized medium: A comprehensive study in the
  Nambu\textendash{}Jona-Lasinio model}},\ }\href
  {https://doi.org/10.1103/PhysRevD.103.094001} {\bibfield  {journal} {\bibinfo
   {journal} {Phys. Rev. D}\ }\textbf {\bibinfo {volume} {103}},\ \bibinfo
  {pages} {094001} (\bibinfo {year} {2021})},\ \Eprint
  {https://arxiv.org/abs/2010.05716} {arXiv:2010.05716 [hep-ph]} \BibitemShut
  {NoStop}%
\bibitem [{\citenamefont {Mei}\ \emph {et~al.}(2022)\citenamefont {Mei},
  \citenamefont {Xia},\ and\ \citenamefont {Mao}}]{Mei:2022dkd}%
  \BibitemOpen
  \bibfield  {author} {\bibinfo {author} {\bibfnamefont {J.}~\bibnamefont
  {Mei}}, \bibinfo {author} {\bibfnamefont {T.}~\bibnamefont {Xia}},\ and\
  \bibinfo {author} {\bibfnamefont {S.}~\bibnamefont {Mao}},\ }\bibfield
  {title} {\bibinfo {title} {{Mass spectra of neutral mesons $K_0,\ \pi_0,\
  \eta,\ \eta'$ at finite magnetic field, temperature and baryon chemical
  potential}},\ }\href@noop {} {\  (\bibinfo {year} {2022})},\ \Eprint
  {https://arxiv.org/abs/2212.04778} {arXiv:2212.04778 [hep-ph]} \BibitemShut
  {NoStop}%
\bibitem [{\citenamefont {Mei}\ and\ \citenamefont {Mao}(2020)}]{Mei:2020jzn}%
  \BibitemOpen
  \bibfield  {author} {\bibinfo {author} {\bibfnamefont {J.}~\bibnamefont
  {Mei}}\ and\ \bibinfo {author} {\bibfnamefont {S.}~\bibnamefont {Mao}},\
  }\bibfield  {title} {\bibinfo {title} {{Inverse catalysis effect of the quark
  anomalous magnetic moment to chiral restoration and deconfinement phase
  transitions}},\ }\href {https://doi.org/10.1103/PhysRevD.102.114035}
  {\bibfield  {journal} {\bibinfo  {journal} {Phys. Rev. D}\ }\textbf {\bibinfo
  {volume} {102}},\ \bibinfo {pages} {114035} (\bibinfo {year} {2020})},\
  \Eprint {https://arxiv.org/abs/2008.12123} {arXiv:2008.12123 [hep-ph]}
  \BibitemShut {NoStop}%
\bibitem [{\citenamefont {Islam}\ \emph {et~al.}(2023)\citenamefont {Islam},
  \citenamefont {Ali},\ and\ \citenamefont {Huang}}]{Islam:2023zyo}%
  \BibitemOpen
  \bibfield  {author} {\bibinfo {author} {\bibfnamefont {C.~A.}\ \bibnamefont
  {Islam}}, \bibinfo {author} {\bibfnamefont {M.~S.}\ \bibnamefont {Ali}},\
  and\ \bibinfo {author} {\bibfnamefont {M.}~\bibnamefont {Huang}},\ }\bibfield
   {title} {\bibinfo {title} {{Deciding on the anomalous magnetic moment of
  quarks in a framework of nonlocal NJL model}},\ }\href@noop {} {\  (\bibinfo
  {year} {2023})},\ \Eprint {https://arxiv.org/abs/2302.00696}
  {arXiv:2302.00696 [hep-ph]} \BibitemShut {NoStop}%
\bibitem [{\citenamefont {Das}\ and\ \citenamefont
  {Haque}(2020)}]{Das:2019ehv}%
  \BibitemOpen
  \bibfield  {author} {\bibinfo {author} {\bibfnamefont {A.}~\bibnamefont
  {Das}}\ and\ \bibinfo {author} {\bibfnamefont {N.}~\bibnamefont {Haque}},\
  }\bibfield  {title} {\bibinfo {title} {{Neutral pion mass in the linear sigma
  model coupled to quarks at arbitrary magnetic field}},\ }\href
  {https://doi.org/10.1103/PhysRevD.101.074033} {\bibfield  {journal} {\bibinfo
   {journal} {Phys. Rev. D}\ }\textbf {\bibinfo {volume} {101}},\ \bibinfo
  {pages} {074033} (\bibinfo {year} {2020})},\ \Eprint
  {https://arxiv.org/abs/1908.10323} {arXiv:1908.10323 [hep-ph]} \BibitemShut
  {NoStop}%
\bibitem [{\citenamefont {Schaefer}\ and\ \citenamefont
  {Wambach}(2005)}]{Schaefer:2004en}%
  \BibitemOpen
  \bibfield  {author} {\bibinfo {author} {\bibfnamefont {B.-J.}\ \bibnamefont
  {Schaefer}}\ and\ \bibinfo {author} {\bibfnamefont {J.}~\bibnamefont
  {Wambach}},\ }\bibfield  {title} {\bibinfo {title} {{The Phase diagram of the
  quark meson model}},\ }\href
  {https://doi.org/10.1016/j.nuclphysa.2005.04.012} {\bibfield  {journal}
  {\bibinfo  {journal} {Nucl. Phys. A}\ }\textbf {\bibinfo {volume} {757}},\
  \bibinfo {pages} {479} (\bibinfo {year} {2005})},\ \Eprint
  {https://arxiv.org/abs/nucl-th/0403039} {arXiv:nucl-th/0403039} \BibitemShut
  {NoStop}%
\bibitem [{\citenamefont {Chen}\ \emph {et~al.}(2021)\citenamefont {Chen},
  \citenamefont {Wen},\ and\ \citenamefont {Fu}}]{Chen:2021iuo}%
  \BibitemOpen
  \bibfield  {author} {\bibinfo {author} {\bibfnamefont {Y.-r.}\ \bibnamefont
  {Chen}}, \bibinfo {author} {\bibfnamefont {R.}~\bibnamefont {Wen}},\ and\
  \bibinfo {author} {\bibfnamefont {W.-j.}\ \bibnamefont {Fu}},\ }\bibfield
  {title} {\bibinfo {title} {{Critical behaviors of the O(4) and Z(2)
  symmetries in the QCD phase diagram}},\ }\href
  {https://doi.org/10.1103/PhysRevD.104.054009} {\bibfield  {journal} {\bibinfo
   {journal} {Phys. Rev. D}\ }\textbf {\bibinfo {volume} {104}},\ \bibinfo
  {pages} {054009} (\bibinfo {year} {2021})},\ \Eprint
  {https://arxiv.org/abs/2101.08484} {arXiv:2101.08484 [hep-ph]} \BibitemShut
  {NoStop}%
\bibitem [{\citenamefont {Schaefer}\ and\ \citenamefont
  {Pirner}(1999)}]{Schaefer:1999em}%
  \BibitemOpen
  \bibfield  {author} {\bibinfo {author} {\bibfnamefont {B.-J.}\ \bibnamefont
  {Schaefer}}\ and\ \bibinfo {author} {\bibfnamefont {H.-J.}\ \bibnamefont
  {Pirner}},\ }\bibfield  {title} {\bibinfo {title} {{Renormalization group
  flow and equation of state of quarks and mesons}},\ }\href
  {https://doi.org/10.1016/S0375-9474(99)00409-1} {\bibfield  {journal}
  {\bibinfo  {journal} {Nucl. Phys. A}\ }\textbf {\bibinfo {volume} {660}},\
  \bibinfo {pages} {439} (\bibinfo {year} {1999})},\ \Eprint
  {https://arxiv.org/abs/nucl-th/9903003} {arXiv:nucl-th/9903003} \BibitemShut
  {NoStop}%
\bibitem [{\citenamefont {Herbst}\ \emph {et~al.}(2014)\citenamefont {Herbst},
  \citenamefont {Mitter}, \citenamefont {Pawlowski}, \citenamefont {Schaefer},\
  and\ \citenamefont {Stiele}}]{Herbst:2013ufa}%
  \BibitemOpen
  \bibfield  {author} {\bibinfo {author} {\bibfnamefont {T.~K.}\ \bibnamefont
  {Herbst}}, \bibinfo {author} {\bibfnamefont {M.}~\bibnamefont {Mitter}},
  \bibinfo {author} {\bibfnamefont {J.~M.}\ \bibnamefont {Pawlowski}}, \bibinfo
  {author} {\bibfnamefont {B.-J.}\ \bibnamefont {Schaefer}},\ and\ \bibinfo
  {author} {\bibfnamefont {R.}~\bibnamefont {Stiele}},\ }\bibfield  {title}
  {\bibinfo {title} {{Thermodynamics of QCD at vanishing density}},\ }\href
  {https://doi.org/10.1016/j.physletb.2014.02.045} {\bibfield  {journal}
  {\bibinfo  {journal} {Phys. Lett. B}\ }\textbf {\bibinfo {volume} {731}},\
  \bibinfo {pages} {248} (\bibinfo {year} {2014})},\ \Eprint
  {https://arxiv.org/abs/1308.3621} {arXiv:1308.3621 [hep-ph]} \BibitemShut
  {NoStop}%
\bibitem [{\citenamefont {Wen}\ \emph {et~al.}(2019)\citenamefont {Wen},
  \citenamefont {Huang},\ and\ \citenamefont {Fu}}]{Wen:2018nkn}%
  \BibitemOpen
  \bibfield  {author} {\bibinfo {author} {\bibfnamefont {R.}~\bibnamefont
  {Wen}}, \bibinfo {author} {\bibfnamefont {C.}~\bibnamefont {Huang}},\ and\
  \bibinfo {author} {\bibfnamefont {W.-J.}\ \bibnamefont {Fu}},\ }\bibfield
  {title} {\bibinfo {title} {{Baryon number fluctuations in the 2+1 flavor low
  energy effective model}},\ }\href
  {https://doi.org/10.1103/PhysRevD.99.094019} {\bibfield  {journal} {\bibinfo
  {journal} {Phys. Rev. D}\ }\textbf {\bibinfo {volume} {99}},\ \bibinfo
  {pages} {094019} (\bibinfo {year} {2019})},\ \Eprint
  {https://arxiv.org/abs/1809.04233} {arXiv:1809.04233 [hep-ph]} \BibitemShut
  {NoStop}%
\bibitem [{\citenamefont {Fu}\ \emph {et~al.}(2021)\citenamefont {Fu},
  \citenamefont {Luo}, \citenamefont {Pawlowski}, \citenamefont {Rennecke},
  \citenamefont {Wen},\ and\ \citenamefont {Yin}}]{Fu:2021oaw}%
  \BibitemOpen
  \bibfield  {author} {\bibinfo {author} {\bibfnamefont {W.-j.}\ \bibnamefont
  {Fu}}, \bibinfo {author} {\bibfnamefont {X.}~\bibnamefont {Luo}}, \bibinfo
  {author} {\bibfnamefont {J.~M.}\ \bibnamefont {Pawlowski}}, \bibinfo {author}
  {\bibfnamefont {F.}~\bibnamefont {Rennecke}}, \bibinfo {author}
  {\bibfnamefont {R.}~\bibnamefont {Wen}},\ and\ \bibinfo {author}
  {\bibfnamefont {S.}~\bibnamefont {Yin}},\ }\bibfield  {title} {\bibinfo
  {title} {{Hyper-order baryon number fluctuations at finite temperature and
  density}},\ }\href@noop {} {\  (\bibinfo {year} {2021})},\ \Eprint
  {https://arxiv.org/abs/2101.06035} {arXiv:2101.06035 [hep-ph]} \BibitemShut
  {NoStop}%
\bibitem [{\citenamefont {Hubbard}(1959)}]{Hubbard:1959ub}%
  \BibitemOpen
  \bibfield  {author} {\bibinfo {author} {\bibfnamefont {J.}~\bibnamefont
  {Hubbard}},\ }\bibfield  {title} {\bibinfo {title} {{Calculation of partition
  functions}},\ }\href {https://doi.org/10.1103/PhysRevLett.3.77} {\bibfield
  {journal} {\bibinfo  {journal} {Phys. Rev. Lett.}\ }\textbf {\bibinfo
  {volume} {3}},\ \bibinfo {pages} {77} (\bibinfo {year} {1959})}\BibitemShut
  {NoStop}%
\bibitem [{\citenamefont {Jung}\ \emph {et~al.}(2017)\citenamefont {Jung},
  \citenamefont {Rennecke}, \citenamefont {Tripolt}, \citenamefont {von
  Smekal},\ and\ \citenamefont {Wambach}}]{Jung:2016yxl}%
  \BibitemOpen
  \bibfield  {author} {\bibinfo {author} {\bibfnamefont {C.}~\bibnamefont
  {Jung}}, \bibinfo {author} {\bibfnamefont {F.}~\bibnamefont {Rennecke}},
  \bibinfo {author} {\bibfnamefont {R.-A.}\ \bibnamefont {Tripolt}}, \bibinfo
  {author} {\bibfnamefont {L.}~\bibnamefont {von Smekal}},\ and\ \bibinfo
  {author} {\bibfnamefont {J.}~\bibnamefont {Wambach}},\ }\bibfield  {title}
  {\bibinfo {title} {{In-Medium Spectral Functions of Vector- and Axial-Vector
  Mesons from the Functional Renormalization Group}},\ }\href
  {https://doi.org/10.1103/PhysRevD.95.036020} {\bibfield  {journal} {\bibinfo
  {journal} {Phys. Rev. D}\ }\textbf {\bibinfo {volume} {95}},\ \bibinfo
  {pages} {036020} (\bibinfo {year} {2017})},\ \Eprint
  {https://arxiv.org/abs/1610.08754} {arXiv:1610.08754 [hep-ph]} \BibitemShut
  {NoStop}%
\bibitem [{\citenamefont {Dupuis}\ \emph {et~al.}(2021)\citenamefont {Dupuis},
  \citenamefont {Canet}, \citenamefont {Eichhorn}, \citenamefont {Metzner},
  \citenamefont {Pawlowski}, \citenamefont {Tissier},\ and\ \citenamefont
  {Wschebor}}]{Dupuis:2020fhh}%
  \BibitemOpen
  \bibfield  {author} {\bibinfo {author} {\bibfnamefont {N.}~\bibnamefont
  {Dupuis}}, \bibinfo {author} {\bibfnamefont {L.}~\bibnamefont {Canet}},
  \bibinfo {author} {\bibfnamefont {A.}~\bibnamefont {Eichhorn}}, \bibinfo
  {author} {\bibfnamefont {W.}~\bibnamefont {Metzner}}, \bibinfo {author}
  {\bibfnamefont {J.~M.}\ \bibnamefont {Pawlowski}}, \bibinfo {author}
  {\bibfnamefont {M.}~\bibnamefont {Tissier}},\ and\ \bibinfo {author}
  {\bibfnamefont {N.}~\bibnamefont {Wschebor}},\ }\bibfield  {title} {\bibinfo
  {title} {{The nonperturbative functional renormalization group and its
  applications}},\ }\href {https://doi.org/10.1016/j.physrep.2021.01.001}
  {\bibfield  {journal} {\bibinfo  {journal} {Phys. Rept.}\ }\textbf {\bibinfo
  {volume} {910}},\ \bibinfo {pages} {1} (\bibinfo {year} {2021})},\ \Eprint
  {https://arxiv.org/abs/2006.04853} {arXiv:2006.04853 [cond-mat.stat-mech]}
  \BibitemShut {NoStop}%
\bibitem [{\citenamefont {Pawlowski}(2007)}]{Pawlowski:2005xe}%
  \BibitemOpen
  \bibfield  {author} {\bibinfo {author} {\bibfnamefont {J.~M.}\ \bibnamefont
  {Pawlowski}},\ }\bibfield  {title} {\bibinfo {title} {{Aspects of the
  functional renormalisation group}},\ }\href
  {https://doi.org/10.1016/j.aop.2007.01.007} {\bibfield  {journal} {\bibinfo
  {journal} {Annals Phys.}\ }\textbf {\bibinfo {volume} {322}},\ \bibinfo
  {pages} {2831} (\bibinfo {year} {2007})},\ \Eprint
  {https://arxiv.org/abs/hep-th/0512261} {arXiv:hep-th/0512261} \BibitemShut
  {NoStop}%
\bibitem [{\citenamefont {Fu}\ \emph {et~al.}(2020)\citenamefont {Fu},
  \citenamefont {Pawlowski},\ and\ \citenamefont {Rennecke}}]{Fu:2019hdw}%
  \BibitemOpen
  \bibfield  {author} {\bibinfo {author} {\bibfnamefont {W.-j.}\ \bibnamefont
  {Fu}}, \bibinfo {author} {\bibfnamefont {J.~M.}\ \bibnamefont {Pawlowski}},\
  and\ \bibinfo {author} {\bibfnamefont {F.}~\bibnamefont {Rennecke}},\
  }\bibfield  {title} {\bibinfo {title} {{QCD phase structure at finite
  temperature and density}},\ }\href
  {https://doi.org/10.1103/PhysRevD.101.054032} {\bibfield  {journal} {\bibinfo
   {journal} {Phys. Rev. D}\ }\textbf {\bibinfo {volume} {101}},\ \bibinfo
  {pages} {054032} (\bibinfo {year} {2020})},\ \Eprint
  {https://arxiv.org/abs/1909.02991} {arXiv:1909.02991 [hep-ph]} \BibitemShut
  {NoStop}%
\bibitem [{\citenamefont {Braun}\ \emph
  {et~al.}(2016{\natexlab{b}})\citenamefont {Braun}, \citenamefont {Fister},
  \citenamefont {Pawlowski},\ and\ \citenamefont {Rennecke}}]{Braun:2014ata}%
  \BibitemOpen
  \bibfield  {author} {\bibinfo {author} {\bibfnamefont {J.}~\bibnamefont
  {Braun}}, \bibinfo {author} {\bibfnamefont {L.}~\bibnamefont {Fister}},
  \bibinfo {author} {\bibfnamefont {J.~M.}\ \bibnamefont {Pawlowski}},\ and\
  \bibinfo {author} {\bibfnamefont {F.}~\bibnamefont {Rennecke}},\ }\bibfield
  {title} {\bibinfo {title} {{From Quarks and Gluons to Hadrons: Chiral
  Symmetry Breaking in Dynamical QCD}},\ }\href
  {https://doi.org/10.1103/PhysRevD.94.034016} {\bibfield  {journal} {\bibinfo
  {journal} {Phys. Rev. D}\ }\textbf {\bibinfo {volume} {94}},\ \bibinfo
  {pages} {034016} (\bibinfo {year} {2016}{\natexlab{b}})},\ \Eprint
  {https://arxiv.org/abs/1412.1045} {arXiv:1412.1045 [hep-ph]} \BibitemShut
  {NoStop}%
\bibitem [{\citenamefont {Helmboldt}\ \emph {et~al.}(2015)\citenamefont
  {Helmboldt}, \citenamefont {Pawlowski},\ and\ \citenamefont
  {Strodthoff}}]{Helmboldt:2014iya}%
  \BibitemOpen
  \bibfield  {author} {\bibinfo {author} {\bibfnamefont {A.~J.}\ \bibnamefont
  {Helmboldt}}, \bibinfo {author} {\bibfnamefont {J.~M.}\ \bibnamefont
  {Pawlowski}},\ and\ \bibinfo {author} {\bibfnamefont {N.}~\bibnamefont
  {Strodthoff}},\ }\bibfield  {title} {\bibinfo {title} {{Towards quantitative
  precision in the chiral crossover: masses and fluctuation scales}},\ }\href
  {https://doi.org/10.1103/PhysRevD.91.054010} {\bibfield  {journal} {\bibinfo
  {journal} {Phys. Rev. D}\ }\textbf {\bibinfo {volume} {91}},\ \bibinfo
  {pages} {054010} (\bibinfo {year} {2015})},\ \Eprint
  {https://arxiv.org/abs/1409.8414} {arXiv:1409.8414 [hep-ph]} \BibitemShut
  {NoStop}%
\bibitem [{\citenamefont {Andersen}\ \emph {et~al.}(2014)\citenamefont
  {Andersen}, \citenamefont {Naylor},\ and\ \citenamefont
  {Tranberg}}]{Andersen:2013swa}%
  \BibitemOpen
  \bibfield  {author} {\bibinfo {author} {\bibfnamefont {J.~O.}\ \bibnamefont
  {Andersen}}, \bibinfo {author} {\bibfnamefont {W.~R.}\ \bibnamefont
  {Naylor}},\ and\ \bibinfo {author} {\bibfnamefont {A.}~\bibnamefont
  {Tranberg}},\ }\bibfield  {title} {\bibinfo {title} {{Chiral and
  deconfinement transitions in a magnetic background using the functional
  renormalization group with the Polyakov loop}},\ }\href
  {https://doi.org/10.1007/JHEP04(2014)187} {\bibfield  {journal} {\bibinfo
  {journal} {JHEP}\ }\textbf {\bibinfo {volume} {04}},\ \bibinfo {pages}
  {187}},\ \Eprint {https://arxiv.org/abs/1311.2093} {arXiv:1311.2093 [hep-ph]}
  \BibitemShut {NoStop}%
\bibitem [{\citenamefont {Wetterich}(1993)}]{Wetterich:1992yh}%
  \BibitemOpen
  \bibfield  {author} {\bibinfo {author} {\bibfnamefont {C.}~\bibnamefont
  {Wetterich}},\ }\bibfield  {title} {\bibinfo {title} {{Exact evolution
  equation for the effective potential}},\ }\href
  {https://doi.org/10.1016/0370-2693(93)90726-X} {\bibfield  {journal}
  {\bibinfo  {journal} {Phys. Lett. B}\ }\textbf {\bibinfo {volume} {301}},\
  \bibinfo {pages} {90} (\bibinfo {year} {1993})},\ \Eprint
  {https://arxiv.org/abs/1710.05815} {arXiv:1710.05815 [hep-th]} \BibitemShut
  {NoStop}%
\bibitem [{\citenamefont {Yin}\ \emph {et~al.}(2019)\citenamefont {Yin},
  \citenamefont {Wen},\ and\ \citenamefont {Fu}}]{Yin:2019ebz}%
  \BibitemOpen
  \bibfield  {author} {\bibinfo {author} {\bibfnamefont {S.}~\bibnamefont
  {Yin}}, \bibinfo {author} {\bibfnamefont {R.}~\bibnamefont {Wen}},\ and\
  \bibinfo {author} {\bibfnamefont {W.-j.}\ \bibnamefont {Fu}},\ }\bibfield
  {title} {\bibinfo {title} {{Mesonic dynamics and the QCD phase transition}},\
  }\href {https://doi.org/10.1103/PhysRevD.100.094029} {\bibfield  {journal}
  {\bibinfo  {journal} {Phys. Rev. D}\ }\textbf {\bibinfo {volume} {100}},\
  \bibinfo {pages} {094029} (\bibinfo {year} {2019})},\ \Eprint
  {https://arxiv.org/abs/1907.10262} {arXiv:1907.10262 [hep-ph]} \BibitemShut
  {NoStop}%
\bibitem [{\citenamefont {Cao}(2021)}]{Cao:2021rwx}%
  \BibitemOpen
  \bibfield  {author} {\bibinfo {author} {\bibfnamefont {G.}~\bibnamefont
  {Cao}},\ }\bibfield  {title} {\bibinfo {title} {{Recent progresses on QCD
  phases in a strong magnetic field: views from Nambu\textendash{}Jona-Lasinio
  model}},\ }\href {https://doi.org/10.1140/epja/s10050-021-00570-0} {\bibfield
   {journal} {\bibinfo  {journal} {Eur. Phys. J. A}\ }\textbf {\bibinfo
  {volume} {57}},\ \bibinfo {pages} {264} (\bibinfo {year} {2021})},\ \Eprint
  {https://arxiv.org/abs/2103.00456} {arXiv:2103.00456 [hep-ph]} \BibitemShut
  {NoStop}%
\bibitem [{\citenamefont {Schwinger}(1951)}]{Schwinger:1951nm}%
  \BibitemOpen
  \bibfield  {author} {\bibinfo {author} {\bibfnamefont {J.~S.}\ \bibnamefont
  {Schwinger}},\ }\bibfield  {title} {\bibinfo {title} {{On gauge invariance
  and vacuum polarization}},\ }\href {https://doi.org/10.1103/PhysRev.82.664}
  {\bibfield  {journal} {\bibinfo  {journal} {Phys. Rev.}\ }\textbf {\bibinfo
  {volume} {82}},\ \bibinfo {pages} {664} (\bibinfo {year} {1951})}\BibitemShut
  {NoStop}%
\bibitem [{\citenamefont {Mao}(2019)}]{Mao:2018dqe}%
  \BibitemOpen
  \bibfield  {author} {\bibinfo {author} {\bibfnamefont {S.}~\bibnamefont
  {Mao}},\ }\bibfield  {title} {\bibinfo {title} {{Pions in magnetic field at
  finite temperature}},\ }\href {https://doi.org/10.1103/PhysRevD.99.056005}
  {\bibfield  {journal} {\bibinfo  {journal} {Phys. Rev. D}\ }\textbf {\bibinfo
  {volume} {99}},\ \bibinfo {pages} {056005} (\bibinfo {year} {2019})},\
  \Eprint {https://arxiv.org/abs/1808.10242} {arXiv:1808.10242 [nucl-th]}
  \BibitemShut {NoStop}%
\bibitem [{\citenamefont {Chyi}\ \emph {et~al.}(2000)\citenamefont {Chyi},
  \citenamefont {Hwang}, \citenamefont {Kao}, \citenamefont {Lin},
  \citenamefont {Ng},\ and\ \citenamefont {Tseng}}]{Chyi:1999fc}%
  \BibitemOpen
  \bibfield  {author} {\bibinfo {author} {\bibfnamefont {T.-K.}\ \bibnamefont
  {Chyi}}, \bibinfo {author} {\bibfnamefont {C.-W.}\ \bibnamefont {Hwang}},
  \bibinfo {author} {\bibfnamefont {W.~F.}\ \bibnamefont {Kao}}, \bibinfo
  {author} {\bibfnamefont {G.-L.}\ \bibnamefont {Lin}}, \bibinfo {author}
  {\bibfnamefont {K.-W.}\ \bibnamefont {Ng}},\ and\ \bibinfo {author}
  {\bibfnamefont {J.-J.}\ \bibnamefont {Tseng}},\ }\bibfield  {title} {\bibinfo
  {title} {{The weak field expansion for processes in a homogeneous background
  magnetic field}},\ }\href {https://doi.org/10.1103/PhysRevD.62.105014}
  {\bibfield  {journal} {\bibinfo  {journal} {Phys. Rev. D}\ }\textbf {\bibinfo
  {volume} {62}},\ \bibinfo {pages} {105014} (\bibinfo {year} {2000})},\
  \Eprint {https://arxiv.org/abs/hep-th/9912134} {arXiv:hep-th/9912134}
  \BibitemShut {NoStop}%
\bibitem [{\citenamefont {Ayala}\ \emph {et~al.}(2015)\citenamefont {Ayala},
  \citenamefont {Loewe},\ and\ \citenamefont {Zamora}}]{Ayala:2014gwa}%
  \BibitemOpen
  \bibfield  {author} {\bibinfo {author} {\bibfnamefont {A.}~\bibnamefont
  {Ayala}}, \bibinfo {author} {\bibfnamefont {M.}~\bibnamefont {Loewe}},\ and\
  \bibinfo {author} {\bibfnamefont {R.}~\bibnamefont {Zamora}},\ }\bibfield
  {title} {\bibinfo {title} {{Inverse magnetic catalysis in the linear sigma
  model with quarks}},\ }\href {https://doi.org/10.1103/PhysRevD.91.016002}
  {\bibfield  {journal} {\bibinfo  {journal} {Phys. Rev. D}\ }\textbf {\bibinfo
  {volume} {91}},\ \bibinfo {pages} {016002} (\bibinfo {year} {2015})},\
  \Eprint {https://arxiv.org/abs/1406.7408} {arXiv:1406.7408 [hep-ph]}
  \BibitemShut {NoStop}%
\bibitem [{\citenamefont {Ayala}\ \emph {et~al.}(2005)\citenamefont {Ayala},
  \citenamefont {Sanchez}, \citenamefont {Piccinelli},\ and\ \citenamefont
  {Sahu}}]{Ayala:2004dx}%
  \BibitemOpen
  \bibfield  {author} {\bibinfo {author} {\bibfnamefont {A.}~\bibnamefont
  {Ayala}}, \bibinfo {author} {\bibfnamefont {A.}~\bibnamefont {Sanchez}},
  \bibinfo {author} {\bibfnamefont {G.}~\bibnamefont {Piccinelli}},\ and\
  \bibinfo {author} {\bibfnamefont {S.}~\bibnamefont {Sahu}},\ }\bibfield
  {title} {\bibinfo {title} {{Effective potential at finite temperature in a
  constant magnetic field. I. Ring diagrams in a scalar theory}},\ }\href
  {https://doi.org/10.1103/PhysRevD.71.023004} {\bibfield  {journal} {\bibinfo
  {journal} {Phys. Rev. D}\ }\textbf {\bibinfo {volume} {71}},\ \bibinfo
  {pages} {023004} (\bibinfo {year} {2005})},\ \Eprint
  {https://arxiv.org/abs/hep-ph/0412135} {arXiv:hep-ph/0412135} \BibitemShut
  {NoStop}%
\bibitem [{\citenamefont {Pawlowski}\ and\ \citenamefont
  {Rennecke}(2014)}]{Pawlowski:2014zaa}%
  \BibitemOpen
  \bibfield  {author} {\bibinfo {author} {\bibfnamefont {J.~M.}\ \bibnamefont
  {Pawlowski}}\ and\ \bibinfo {author} {\bibfnamefont {F.}~\bibnamefont
  {Rennecke}},\ }\bibfield  {title} {\bibinfo {title} {{Higher order
  quark-mesonic scattering processes and the phase structure of QCD}},\ }\href
  {https://doi.org/10.1103/PhysRevD.90.076002} {\bibfield  {journal} {\bibinfo
  {journal} {Phys. Rev. D}\ }\textbf {\bibinfo {volume} {90}},\ \bibinfo
  {pages} {076002} (\bibinfo {year} {2014})},\ \Eprint
  {https://arxiv.org/abs/1403.1179} {arXiv:1403.1179 [hep-ph]} \BibitemShut
  {NoStop}%
\end{thebibliography}%

\end{document}